\PassOptionsToPackage{unicode}{hyperref}
\PassOptionsToPackage{hyphens}{url}
\PassOptionsToPackage{dvipsnames,svgnames,x11names}{xcolor}
\documentclass[
  11pt,
  a4paper]{article}
\usepackage{xcolor}
\usepackage[margin=2.5cm]{geometry}
\usepackage{amsmath,amssymb}
\usepackage{iftex}
\ifPDFTeX
  \usepackage[T1]{fontenc}
  \usepackage[utf8]{inputenc}
  \usepackage{textcomp} 
\else 
  \usepackage{unicode-math} 
  \defaultfontfeatures{Scale=MatchLowercase}
  \defaultfontfeatures[\rmfamily]{Ligatures=TeX,Scale=1}
\fi
\usepackage{lmodern}
\ifPDFTeX\else
\fi
\IfFileExists{upquote.sty}{\usepackage{upquote}}{}
\IfFileExists{microtype.sty}{
  \usepackage[]{microtype}
  \UseMicrotypeSet[protrusion]{basicmath} 
}{}
\makeatletter
\@ifundefined{KOMAClassName}{
  \IfFileExists{parskip.sty}{%
    \usepackage{parskip}
  }{
    \setlength{\parindent}{0pt}
    \setlength{\parskip}{6pt plus 2pt minus 1pt}}
}{
  \KOMAoptions{parskip=half}}
\makeatother
\makeatletter
\ifx\paragraph\undefined\else
  \let\oldparagraph\paragraph
  \renewcommand{\paragraph}{
    \@ifstar
      \xxxParagraphStar
      \xxxParagraphNoStar
  }
  \newcommand{\xxxParagraphStar}[1]{\oldparagraph*{#1}\mbox{}}
  \newcommand{\xxxParagraphNoStar}[1]{\oldparagraph{#1}\mbox{}}
\fi
\ifx\subparagraph\undefined\else
  \let\oldsubparagraph\subparagraph
  \renewcommand{\subparagraph}{
    \@ifstar
      \xxxSubParagraphStar
      \xxxSubParagraphNoStar
  }
  \newcommand{\xxxSubParagraphStar}[1]{\oldsubparagraph*{#1}\mbox{}}
  \newcommand{\xxxSubParagraphNoStar}[1]{\oldsubparagraph{#1}\mbox{}}
\fi
\makeatother

\usepackage{longtable,booktabs,array}
\usepackage{calc} 
\usepackage{etoolbox}
\makeatletter
\patchcmd\longtable{\par}{\if@noskipsec\mbox{}\fi\par}{}{}
\makeatother
\IfFileExists{footnotehyper.sty}{\usepackage{footnotehyper}}{\usepackage{footnote}}
\makesavenoteenv{longtable}
\usepackage{graphicx}
\makeatletter
\newsavebox\pandoc@box
\newcommand*\pandocbounded[1]{
  \sbox\pandoc@box{#1}%
  \Gscale@div\@tempa{\textheight}{\dimexpr\ht\pandoc@box+\dp\pandoc@box\relax}%
  \Gscale@div\@tempb{\linewidth}{\wd\pandoc@box}%
  \ifdim\@tempb\p@<\@tempa\p@\let\@tempa\@tempb\fi
  \ifdim\@tempa\p@<\p@\scalebox{\@tempa}{\usebox\pandoc@box}%
  \else\usebox{\pandoc@box}%
  \fi%
}
\def\fps@figure{htbp}
\makeatother

\NewDocumentCommand\citeproctext{}{}

\makeatletter
 \let\@cite@ofmt\@firstofone
 \def\@biblabel#1{}
 \def\@cite#1#2{{#1\if@tempswa , #2\fi}}
\makeatother
\newlength{\cslhangindent}
\newlength{\csllabelwidth}
\newenvironment{CSLReferences}[2] 
 {\begin{list}{}{%
  \setlength{\itemindent}{0pt}
  \setlength{\leftmargin}{0pt}
  \setlength{\parsep}{0pt}
  \ifodd #1
   \setlength{\leftmargin}{\cslhangindent}
   \setlength{\itemindent}{-1\cslhangindent}
  \fi
  \setlength{\itemsep}{#2\baselineskip}}}
 {\end{list}}
\usepackage{calc}

\newcommand{\CSLLeftMargin}[1]{\parbox[t]{\csllabelwidth}{\strut#1\strut}}
\newcommand{\CSLRightInline}[1]{\parbox[t]{\linewidth - \csllabelwidth}{\strut#1\strut}}

\providecommand{\tightlist}{%
  \setlength{\itemsep}{0pt}\setlength{\parskip}{0pt}}

\usepackage{microtype}
\usepackage{booktabs}
\usepackage{needspace}

\makeatletter
\@ifpackageloaded{caption}{}{\usepackage{caption}}
\AtBeginDocument{%
\ifdefined\contentsname
  \renewcommand*\contentsname{Table of contents}
\else
  \newcommand\contentsname{Table of contents}
\fi
\ifdefined\listfigurename
  \renewcommand*\listfigurename{List of Figures}
\else
  \newcommand\listfigurename{List of Figures}
\fi
\ifdefined\listtablename
  \renewcommand*\listtablename{List of Tables}
\else
  \newcommand\listtablename{List of Tables}
\fi
\ifdefined\figurename
  \renewcommand*\figurename{Fig.}
\else
  \newcommand\figurename{Fig.}
\fi
\ifdefined\tablename
  \renewcommand*\tablename{Table}
\else
  \newcommand\tablename{Table}
\fi
}
\@ifpackageloaded{float}{}{\usepackage{float}}
\floatstyle{ruled}
\@ifundefined{c@chapter}{\newfloat{codelisting}{h}{lop}}{\newfloat{codelisting}{h}{lop}[chapter]}
\floatname{codelisting}{Listing}

\makeatother
\makeatletter
\@ifpackageloaded{caption}{}{\usepackage{caption}}
\@ifpackageloaded{subcaption}{}{\usepackage{subcaption}}
\makeatother
\usepackage{bookmark}
\IfFileExists{xurl.sty}{\usepackage{xurl}}{} 
\makeatletter
\@ifundefined{xmpquote}{}{}
\makeatother
\hypersetup{
  pdftitle={Reliability Assessment of FRP-RC Columns Using AK-MCS with a Piecewise Two-Branch Approach},
  pdfauthor={Arndt Goldack; Richard Koehler; Ahmed Usama Syed},
  colorlinks=true,
  linkcolor={blue},
  filecolor={Maroon},
  citecolor={blue},
  urlcolor={blue},
  pdfcreator={LaTeX via pandoc}}

\title{Reliability Assessment of FRP-RC Columns Using AK-MCS with a
Piecewise Two-Branch Approach}
\author{Arndt Goldack\thanks{Corresponding author. E-mail: goldack@uni-wuppertal.de. ORCID: 0000-0001-9906-2938.}
\and Richard Koehler\thanks{ORCID: 0009-0008-2430-411X.}
\and Ahmed Usama Syed\thanks{ORCID: 0000-0001-8028-3124.}\\[6pt]
\normalsize Chair of Statics and Dynamics of Structures, University of Wuppertal, Germany}
\date{2026-10-23}
\begin{document}
\maketitle
\begin{abstract}
Slender concrete columns reinforced with fiber-reinforced polymer (FRP)
bars fail at medium to high axial load levels either by concrete
crushing or by loss of stability, while the tensile strains in the
reinforcement remain far below the rupture strain. Each non-linear
finite element (FE) analysis yields only the capacity of the governing
failure mode, and the governing mode changes over a broad zone in which
both capacities are nearly equal. Both characteristics conflict with
active-learning Kriging-assisted Monte Carlo simulation (AK-MCS), which
assumes a single, smooth limit-state function. This paper proposes a
piecewise, two-branch AK-MCS framework in which each failure mode is
represented by its own Kriging model, trained only on the FE analyses in
which that mode governs, and the governing prediction is the minimum of
both. Safeguards against data starvation keep rarely governing branches
usable. After convergence, additional FE analyses are placed based on a
leave-one-out calibration check and on the width of the confidence
interval on the reliability index. A design-point search on the
surrogate and importance sampling at the design points provide the
reliability index even where direct Monte Carlo simulation finds no
failure. For an FRP-reinforced concrete column example with mixed
failure, 79\% and 62\% of the training points lie near the mode
boundary. The reliability index from the design-point search agrees with
the Monte Carlo estimate to within 0.05, and an FE analysis at the
design point confirms the surrogate to within 0.6\%.
\end{abstract}


\subsection*{Nomenclature}\label{nomenclature}
\addcontentsline{toc}{subsection}{Nomenclature}

\Needspace*{8\baselineskip}\noindent\textbf{Limit state and reliability}\par

{\def\LTcaptype{none} 
\begin{longtable}[]{@{}
  >{\raggedright\arraybackslash}p{(\linewidth - 2\tabcolsep) * \real{0.1935}}
  >{\raggedright\arraybackslash}p{(\linewidth - 2\tabcolsep) * \real{0.8065}}@{}}
\toprule\noalign{}
\begin{minipage}[b]{\linewidth}\raggedright
Symbol
\end{minipage} & \begin{minipage}[b]{\linewidth}\raggedright
Meaning
\end{minipage} \\
\midrule\noalign{}
\endhead
\bottomrule\noalign{}
\endlastfoot
\(E(\mathbf{x})\) & action effect \\
\(G_i^{\mp}\) (\(G^-\), \(G^+\)) & pessimistic, optimistic shifted
limit-state function at pool point \(i\),
\(\hat g_\Theta \mp \kappa\hat\sigma_\Theta\) (Section 3.1) \\
\(g(\mathbf{x})\), \(\hat g(\mathbf{x})\) & limit-state function;
Kriging surrogate prediction of limit-state function \\
\(\hat g_\Theta\) & limit-state prediction with \(\Theta_R\),
\(\Theta_E\) applied (Section 2.5) \\
\(N_R\), \(N_{Ed}\) & axial resistance from the FE analysis, used as
\(R\) in the limit-state function; design axial load \\
\(N_{\text{gov}}\), \(N_{\text{mat}}\), \(N_{\text{stab}}\) & idealized
governing capacity,
\(N_{\text{gov}}=\min(N_{\text{stab}},N_{\text{mat}})\) (Section 2.1),
load capacity for material failure, load capacity for stability
failure \\
\(P_f\), \(P_f^+\), \(P_f^-\) & nominal, pessimistic, optimistic failure
probability \\
\(R(\mathbf{x})\), \(\hat R\) & resistance; resistance obtained from the
Kriging prediction, \(\hat R = \hat g + E\) (Section 2.5) \\
\(\mathbf{u}\), \(\mathbf{u}^*\) & standard normal space vector; FORM
design point \\
\(\mathbf{x}\), \(\mathbf{x}_i\), \(\mathbf{x}_j\) & vector of input
variables of \(R\) and \(E\) (here the five surrogate inputs, Section
4.1); pool point \(i\); training point \(j\) \\
\(U(\mathbf{x})\), \(u_{\text{threshold}}\) & active-learning function;
its convergence threshold \\
\(\beta\), \(\beta_{\text{MCS}}\), \(\beta_{\text{MPP}}\) & reliability
index; Monte Carlo pool-based; FORM/MPP-based \\
\(\beta_{\text{indep}}\) & Monte Carlo pool-based \(\beta\) under the
independent-error treatment of the surrogate (Section 3.3) \\
\(\beta_{\text{lower}}\), \(\beta_{\text{upper}}\) & Monte Carlo
pool-based two-sided 95\% confidence bounds on \(\beta\) \\
\(\beta^-_{\text{MPP}}\), \(\beta^+_{\text{MPP}}\) & FORM/MPP search on
the pessimistic, optimistic surfaces \(G^-\), \(G^+\) (Section 3.5) \\
\(\Delta\beta\) & width of the confidence bounds
\(\beta_{\text{lower}}\), \(\beta_{\text{upper}}\) \\
\(\Delta\beta_{\text{FORM}}\) & difference between
\(\beta^-_{\text{MPP}}\) and \(\beta^+_{\text{MPP}}\) \\
\(\delta\beta_{\text{indep}}\) & shift of \(\beta_{\text{indep}}\) from
the nominal \(\beta\) \\
\(\kappa\) & standard normal critical value (\(1.96\) two-sided 95\%;
\(1.645\) one-sided 95\%) \\
\(\hat\sigma\), \(\hat\sigma_\Theta\) & predictive standard deviation of
\(\hat R\); standard deviation of \(\hat g_\Theta\) \\
\(\Theta_R\), \(\Theta_E\) & resistance and action effect model
uncertainty factors \\
\end{longtable}
}

\Needspace*{8\baselineskip}\noindent\textbf{Two-branch surrogate}\par

{\def\LTcaptype{none} 
\begin{longtable}[]{@{}
  >{\raggedright\arraybackslash}p{(\linewidth - 2\tabcolsep) * \real{0.1935}}
  >{\raggedright\arraybackslash}p{(\linewidth - 2\tabcolsep) * \real{0.8065}}@{}}
\toprule\noalign{}
\begin{minipage}[b]{\linewidth}\raggedright
Symbol
\end{minipage} & \begin{minipage}[b]{\linewidth}\raggedright
Meaning
\end{minipage} \\
\midrule\noalign{}
\endhead
\bottomrule\noalign{}
\endlastfoot
\(c_j\), \(c(\mathbf{x})\) & governing mechanism label of FE evaluation
\(j\) (material/stability); failure mode governing at \(\mathbf{x}\)
(Section 2.4) \\
\(D_{\text{mat}}\), \(D_{\text{stab}}\) & training subsets for material
failure and stability failure \\
\(\hat g_{\text{mat}}\), \(\hat g_{\text{stab}}\) & material branch,
stability branch Kriging prediction \\
\(q(\mathbf{u})\) & importance-sampling density \\
\(w(\mathbf{u})\) & importance weight \\
\(s_b\), \(\tilde s_b\), \(\hat s_b\) & win-share of branch \(b\), share
of the candidate pool in which mode \(b\) governs (Section 2.4); direct
count from the min-combination; classifier estimate (Section 2.4.5) \\
\(N_c\) & number of points of the Monte Carlo candidate pool \\
\(p_b(\mathbf{x})\) & classifier probability that mode \(b\) governs at
\(\mathbf{x}\) (Section 2.4.5) \\
\(z_j\) & leave-one-out standardized residual at training point \(j\)
(Section 3.2) \\
\(z_{\text{mode}}\) & normalized distance of a point to the mode
boundary,
\((\hat g_{\text{mat}}-\hat g_{\text{stab}})/(\hat\sigma_{\text{mat}}^2+\hat\sigma_{\text{stab}}^2)^{1/2}\)
(Section 4.3); distinct from \(z_j\) \\
\(\varphi(\mathbf{u})\) & standard normal density \\
\(\hat\sigma_{\text{mat}}\), \(\hat\sigma_{\text{stab}}\) & predictive
standard deviations of \(\hat g_{\text{mat}}\),
\(\hat g_{\text{stab}}\) \\
\end{longtable}
}

\Needspace*{8\baselineskip}\noindent\textbf{Column geometry and materials}\par

{\def\LTcaptype{none} 
\begin{longtable}[]{@{}
  >{\raggedright\arraybackslash}p{(\linewidth - 2\tabcolsep) * \real{0.1935}}
  >{\raggedright\arraybackslash}p{(\linewidth - 2\tabcolsep) * \real{0.8065}}@{}}
\toprule\noalign{}
\begin{minipage}[b]{\linewidth}\raggedright
Symbol
\end{minipage} & \begin{minipage}[b]{\linewidth}\raggedright
Meaning
\end{minipage} \\
\midrule\noalign{}
\endhead
\bottomrule\noalign{}
\endlastfoot
\(A_c\), \(A_g\) & net concrete area,
\(A_c = A_g/(1+\rho_{\text{FRP}})\); gross cross-section area,
\(A_g = b\,h\) \\
\(b\) & cross-section width \\
\(d\) & number of surrogate input variables (here 5) \\
\(d_1\) & concrete cover to the reinforcement centroid \\
\(E_c\) & concrete initial tangent modulus of the JCSS stress--strain
relation (short-term, scaled by \(k_E\)) \\
\(E_{\text{FRP}}\) & FRP bar elastic modulus \\
\(e_0\) & first-order eccentricity \\
\(f_c\) & sampled concrete compressive strength \\
\(f_{ck}\), \(f_{cm}\), \(f_{cd}\), \(f_{cm,r}\) & characteristic, mean
(\(f_{ck}+8\)), design concrete compressive strength; mean compressive
strength used as the Kriging feature (Section 4.1) \\
\(G\), \(G_m\) & dead load; its back-calculated mean value \\
\(h\) & cross-section depth \\
\(k\) & curve-shape factor, \(k=E_c\varepsilon_s/f_c\) \\
\(k_E\) & scaling factor on the modulus of elasticity \(E_c\) of the
concrete model of JCSS {[}1{]} (0.9, 1.0, 1.1); not part of the JCSS
model \\
\(L\) & column length \\
\(M_{01}\), \(M_{02}\) & first-order end moments at column top and
bottom \\
\(M_{Ed}\) & first-order design moment, \(M_{Ed} = N_{Ed}\,e_0\) \\
\(N_e\) & acting axial force, \(N_e = G + Q\) \\
\(Q\), \(Q_m\) & live load; its back-calculated mean value \\
\(Y_1\) & JCSS model-uncertainty variable for the in-situ strength
reduction; not applied in this work \\
\(Y_3\), \(Y_4\) & JCSS model-uncertainty variables for \(E_c\) and for
the ultimate strain \(\varepsilon_u\) \\
\(z_c\) & depth of the strain limit \(\varepsilon_s\) in a fully
compressed section (Section 4.2) \\
\(\alpha_{cc}\) & coefficient for long-term effects on compressive
strength \\
\(\varepsilon_s\), \(\varepsilon_u\) & JCSS reference strain; ultimate
strain \\
\(\varepsilon_{tud}\) & FRP ultimate design tensile strain \\
\(\gamma_c\) & partial factor for concrete,
\(f_{cd}=\alpha_{cc}f_{ck}/\gamma_c\) \\
\(\gamma_g\), \(\gamma_q\) & partial safety factors (dead/live) \\
\(\lambda\) & slenderness ratio \\
\(\lambda_{\text{JCSS}}\) & JCSS in-situ strength exponent,
\(f_c=f_{c0}^{\lambda_{\text{JCSS}}}Y_1\); not applied in this work \\
\(\mu_{Ed}\) & non-dimensional first-order design moment,
\(\mu_{Ed} = M_{Ed}/(A_c\,h\,f_{cd})\) (from the \(N\)--\(M\)
interaction diagrams) \\
\(\mu_g\), \(\mu_q\) & load-share factors (dead/live) \\
\(\nu\) & design axial load ratio, \(\nu = N_{Ed}/(A_c f_{cd})\) \\
\(\omega'\) & mechanical reinforcement ratio,
\(E_{\text{FRP}}\rho_{\text{FRP}}/(100f_{cd})\) \\
\(\rho_{\text{FRP}}\) & FRP reinforcement ratio \\
\(f_c\), \(E_c\), \(E_{\text{FRP}}\), \(N_e\), \(\varepsilon_u\) & the
five-variable Kriging feature set (Section 4.1) \\
\end{longtable}
}

\subsection{1 Introduction}\label{introduction}

Slender concrete columns reinforced with fiber-reinforced polymer (FRP)
bars (FRP-RC columns) pose two challenges for a reliability assessment.
Their load-carrying capacity can be obtained from a materially and
geometrically non-linear finite-element (FE) analysis, and the governing
failure mode is not fixed in advance but depends on the slenderness
ratio (\(\lambda\)), the reinforcement ratio (\(\rho_{\text{FRP}}\)),
the first-order eccentricity (\(e_0\)) and the axial load ratio
(\(\nu\)). At medium to high axial load levels, a column fails either by
concrete crushing (material failure) or by second-order instability
(stability failure). A single FE run returns only the capacity of the
governing mode, never both mode-specific limit-state values.

For designing slender FRP-RC columns, Syed and Goldack {[}2{]} and
Goldack et al. {[}3{]} introduced design equations based on the nominal
curvature method, proposed for the revised DAfStb guideline for
non-metallic reinforcement {[}4{]}. Syed et al. {[}5{]} provide
universal normalized \(N\)--\(M\) interaction diagrams based on these
equations, which were regressed on second-order analyses of 26,880
FRP-RC columns with the direct curvature method. EC2 {[}6{]} applies a
semi-probabilistic safety concept calibrated to a target reliability,
and the forthcoming EC0 {[}7{]} and D375 {[}8{]} introduce a framework
for fully probabilistic verification. For slender columns with steel
reinforcement, the reliability index has been analyzed, e.g.~by Graubner
and Six {[}9{]}. The design equations for slender FRP-RC columns,
however, are neither derived from nor checked against a target
reliability index. It remains unclear whether they provide a sufficient
and consistent level of reliability across their range of validity.

Answering this question requires reliability analyses over a broad range
of design parameters. The limit-state function used throughout this work
is

\begin{equation}\protect\phantomsection\label{eq-limitstate}{g(\mathbf{x}) = \Theta_R R(\mathbf{x}) - \Theta_E E(\mathbf{x}) \leq 0}\end{equation}

Here, \(\mathbf{x}\) is the vector of input variables on which the
resistance \(R\) and the action effect \(E\) depend. \(\Theta_R\) is the
resistance model uncertainty, and \(\Theta_E\) is the model uncertainty
of the action effect, applied to section forces rather than to loads.
All four quantities are random. Indexed vectors denote individual
points, \(\mathbf{x}_i\) a point of the Monte Carlo pool and
\(\mathbf{x}_j\) a training point. A failure probability of \(10^{-6}\)
(\(\beta \approx 4.75\)) requires about \(1.6\cdot10^9\) samples for a
coefficient of variation of \(P_f\) of 2.5\%, which is not feasible by
direct Monte Carlo simulation (MCS) with a non-linear FE model.
Kriging-based active learning (AK-MCS) {[}10,11{]} is therefore used.
Unlike artificial neural networks, Kriging provides a predictive
standard deviation \(\hat\sigma(\mathbf{x})\) at every point, which
drives the learning function and yields a confidence interval on the
reliability index without an additional model.

\subsubsection{1.1 Related work and identified
gap}\label{related-work-and-identified-gap}

Two properties of the present problem limit the use of existing
Kriging-based reliability methods for multiple failure modes.

\begin{enumerate}
\def\labelenumi{\arabic{enumi}.}
\item
  \textbf{Only one governing result}\\
  A single non-linear FE analysis returns only the capacity of the
  governing failure mode, a censored observation. Which mode is censored
  depends systematically on the sample itself, not on an independent
  random process.
\item
  \textbf{Broad and degenerate transition zone}\\
  In critical configurations, up to about 93\% of the training points
  lie within one predictive standard deviation of the mode boundary
  (Section 4.3).
\end{enumerate}

A comprehensive benchmark by Moustapha et al. {[}12{]} compared 39
active-learning strategies on 20 reliability problems. All problems are
posed as a single limit-state function. Even a series system enters as
the minimum over its failure modes and is approximated by one surrogate,
which is the single-model treatment whose difficulties at the kink
motivate the present work (Section 2.1). The authors name multiple
limit-state functions as an aspect still to be investigated.

Nearly all reviewed system-reliability extensions of AK-MCS, such as
EGRA for systems {[}13{]}, AK-SYS {[}14{]}, AK-SYSi {[}15{]} and the
methods of Xiao et al. {[}16{]}, Yuan et al. {[}17{]} and Yang et al.
{[}18,19{]}, assume that every failure mode can be evaluated
independently per sample. The same holds for closed-form reliability
studies on non-metallic reinforcement. Rempel {[}20{]} treated
reinforcement rupture and concrete crushing of textile-reinforced
flexural members as a series system with separate equations for both
capacities. AK-SYS evaluates all modes only for its initial design of
experiments (DoE). Its refinement then evaluates a single mode at each
new point, but this mode is selected by the learning function and must
be computable in isolation {[}14,15,17{]}. A non-linear FE analysis of a
slender column returns only the physically governing mode, so the mode
cannot be selected.

Further extensions resize the candidate pool {[}18{]}, target small
failure probabilities with importance sampling at multiple most probable
points (MPPs) found by a multimodal search {[}19{]}, or replace the pool
by system subset simulation {[}21{]}. All of them keep one Kriging model
per failure mode under the same assumption. With only two branches, the
present approach instead enumerates a small, fixed set of design points
directly (Section 3.6).

Explicit classification of the mode boundary with support vector
machines (SVM) {[}22,23{]} was also examined. In a preliminary study on
the configuration of the example column with the most degenerate mode
boundary, the classifier boundary stabilized with more samples, but it
placed many training points in regions irrelevant for the failure
probability. The confidence interval on \(\beta\) therefore decreased
only marginally. No reviewed method resolves both properties together.

This paper closes this gap with a two-branch Kriging surrogate trained
only on the limit-state values obtained from the capacities that the FE
analyses actually return. Each branch is fitted only where its mode
governs, and the governing prediction is the smaller of the two (Section
2.3). In addition, the first-order reliability method (FORM) is extended
to the two-branch Kriging limit-state surface, including bounds on
\(\beta\) obtained by shifting the surface by \(\pm\kappa\hat\sigma\)
locally at the design point.

\subsubsection{1.2 Contribution and
outline}\label{contribution-and-outline}

This paper introduces a piecewise, two-branch AK-MCS framework for the
two competing failure modes of slender FRP-RC columns at medium to high
axial load levels. First, one Kriging model per failure mode is trained
only on the analyses in which its mode governs. Second, a refinement
after convergence adds training points based on leave-one-out checks and
on the width of the confidence interval on \(\beta\). Third, FORM design
points are identified on the surrogate and on its shifted limit-state
surfaces, giving local confidence bounds on \(\beta\). Finally,
importance sampling with one center per branch resolves small failure
probabilities for both modes. The approach is demonstrated on a
representative column example.

Section 2 introduces the two-branch AK-MCS method. Section 3 derives
confidence bounds on \(\beta\), checks their calibration, extends the
FORM/MPP search to the two-branch surface and adds importance sampling.
Section 4 applies the framework to the column example, including the FE
model. Section 5 discusses the results, and Section 6 concludes.

\subsection{2 Piecewise (Two-Branch)
AK-MCS}\label{piecewise-two-branch-ak-mcs}

\subsubsection{2.1 Motivation}\label{motivation}

Standard AK-MCS {[}10{]} fits a single, stationary Gaussian process (GP)
to the limit-state function and presupposes a continuous response. It
has been applied successfully to conventional reinforced-concrete
members such as bridge girders and columns {[}24{]}, where one failure
mechanism governs throughout. In slender FRP-RC columns, the resistance
is governed by whichever of two distinct mechanisms is triggered first,
which produces a kink at the transition.

\begin{figure}[htbp]

\centering{

\includegraphics[width=0.6\linewidth,height=\textheight,keepaspectratio]{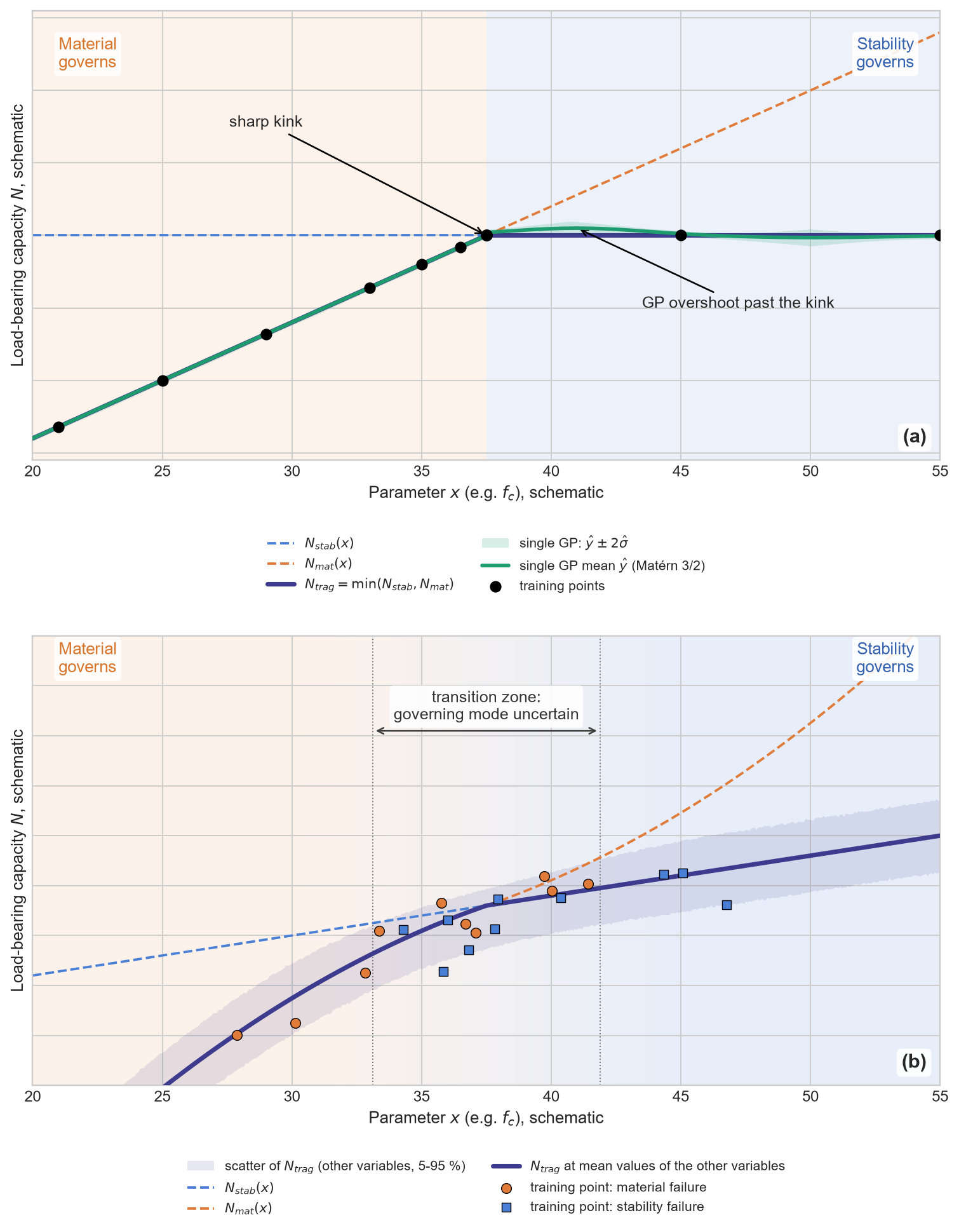}

}

\caption{\label{fig-kink}Two-branch min-concept, schematic. (a)
Idealized sharp kink with single-GP overshoot. (b) Shallow crossing and
broad, blurred transition, as found for the FRP-RC column.}

\end{figure}

The governing capacity can be idealized as
\(N_{\text{gov}}(x) = \min\big(N_{\text{stab}}(x), N_{\text{mat}}(x)\big)\),
each branch smooth but the minimum non-differentiable at the crossing
(Fig.~\ref{fig-kink} a, dark line). Standard covariance kernels (Matérn
3/2, Matérn 5/2, squared exponential) assume a smoothness that this kink
violates. With sparse training data near the kink, a single GP can
overshoot the true function just past the kink (Fig.~\ref{fig-kink} a,
green line). A two-branch surrogate (Section 2.3) avoids this by fitting
the two mechanisms separately.

For the FRP-RC column, the transition is not a sharp kink at a single
point (Fig.~\ref{fig-kink} b). The two capacities \(N_{\text{mat}}\) and
\(N_{\text{stab}}\) intersect at a shallow angle. Because the governing
capacity also depends on the remaining random variables, the observed
failure mode alternates between neighboring training points over a broad
zone. In the example configurations, 79\% and 62\% of the training
points lie within one predictive standard deviation of the mode boundary
(Section 4.3). A companion paper in preparation investigates this
systematically.

\subsubsection{2.2 Standard AK-MCS}\label{standard-ak-mcs}

AK-MCS {[}10{]} trains a Kriging surrogate \(\hat g(\mathbf{x})\) of the
limit-state function on a small initial design, about a dozen points
drawn from the Monte Carlo population, and refines it iteratively over
this population. At each iteration, the point that minimizes the
learning function
\(U(\mathbf{x}) = |\hat g(\mathbf{x})|/\hat\sigma(\mathbf{x})\) is
evaluated and added to the training set. The loop stops once
\(\min U \geq 2\). The failure probability \(P_f\) is then estimated on
the same population, which is enlarged if its coefficient of variation
is too high. In this study, the initial design consists of 15--20
columns from a Latin hypercube design, the candidate pool has
\(10^5\)--\(10^6\) points, the stricter threshold
\(u_{\text{threshold}} = 3\) is used, and \(P_f\) and \(\beta\) are
evaluated by MCS with \(10^7\) samples on the refined surrogate. The
\(U\)-criterion remains one of the most efficient learning criteria
{[}25{]} but presupposes a single continuous failure mechanism.

\subsubsection{2.3 Two-branch surrogate formulation and parallel
prediction}\label{two-branch-surrogate-formulation-and-parallel-prediction}

The initial DoE is drawn over the whole input space and evaluated by FE
before any branch split exists. Each FE evaluation records a
governing-mechanism label \(c_j\in\{\text{mat},\text{stab}\}\) (Section
4.2), which partitions the training set into disjoint subsets
\(D_{\text{mat}}\) and \(D_{\text{stab}}\). On these subsets, two
independent GPs with a Matérn 3/2 kernel predict the limit-state
function for material failure, \(\hat g_{\text{mat}}(\mathbf{x})\), and
for stability failure, \(\hat g_{\text{stab}}(\mathbf{x})\), each with
its own predictive standard deviation. Since the column fails as soon as
either mechanism reaches its limit state, the combined prediction is the
minimum,

\begin{equation}\protect\phantomsection\label{eq-grenzfl}{\hat g(\mathbf{x}) = \min\big(\hat g_{\text{mat}}(\mathbf{x}), \hat g_{\text{stab}}(\mathbf{x})\big) 
}\end{equation}

and its predictive standard deviation \(\hat\sigma(\mathbf{x})\) is that
of the governing branch. This is the series-system combination of AK-SYS
{[}14{]}, but every FE run here yields exactly one label, so the
branches never share training data. The \(U\)-criterion is formed from
the combined prediction of Eq. (\ref{eq-grenzfl}) rather than per branch
as in the composite strategy of AK-SYS. If a branch has fewer than two
points at termination, it remains unfitted, and the model reduces to a
single-branch surrogate.

\subsubsection{2.4 Enhancements for data-scarce
branches}\label{enhancements-for-data-scarce-branches}

Splitting the surrogate introduces a problem that a single-model AK-MCS
does not have. Individual branches can be severely data-scarce,
especially early in the run. Five mechanisms, none with a counterpart in
{[}10{]}, address this problem and are presented below. Three of them,
the minimum-population rule, the fill-up rule and the classifier, rely
on a branch's \textbf{win-share} \(s_b\), the fraction of the Monte
Carlo candidate pool in which failure mode
\(b\in\{\text{mat},\text{stab}\}\) governs. The win-share decides
whether the fit quality of a branch matters. A branch governing 0.3\% of
the pool can be badly fitted without affecting \(P_f\), whereas a branch
governing 50\% cannot. With \(c(\mathbf{x}_i)\) the failure mode
governing at the point \(\mathbf{x}_i\) of a candidate pool of \(N_c\)
points, the win-share is

\begin{equation}\protect\phantomsection\label{eq-winshare-def}{s_b = \frac{1}{N_c}\sum_{i=1}^{N_c}\mathbb{1}\big[c(\mathbf{x}_i) = b\big],\qquad b\in\{\text{mat},\text{stab}\},}\end{equation}

where \(\mathbb{1}[\cdot]\) is the indicator function. Since
\(c(\mathbf{x}_i)\) is known only at the FE-evaluated points, \(s_b\)
must be estimated (Section 2.4.5).

\paragraph{2.4.1 Extrapolation safeguard}\label{extrapolation-safeguard}

Each branch is restricted to an axis-aligned bounding box over the
ranges of its training points in the five surrogate input variables
(Section 4.1), with each limit extended by 10\% of the range. A branch
trained on few, narrowly distributed samples may otherwise extrapolate
beyond its training region and spuriously become the minimum. Its
prediction is therefore set to \(+\infty\) outside its box before the
minimum is taken. If both branches are outside their boxes, the branch
with the smaller \(|\hat g|\) is used. The box can still contain corner
regions without training points, for example a high \(f_c\) combined
with a low \(E_c\), so the safeguard suppresses gross extrapolation but
not every extrapolation. Conversely, a branch with few, widely scattered
points can span a large box and govern \(\beta_{\text{lower}}\) (Section
3.1) through its large \(\hat\sigma\). Excluding such a branch below a
minimum point count removes this effect, but this was not required for
the configurations of this paper.

\paragraph{2.4.2 Minimum population and relevance
threshold}\label{minimum-population-and-relevance-threshold}

If the governing branch has fewer than a minimum number of points (here
10) and its win-share exceeds a relevance threshold of 1\%, \(U\) is set
to zero for all candidate points in its region, so that the next
training point is taken there. If this forcing adds no new point to the
branch within 20 refits, it is switched off, and the branch is treated
as irrelevant. The relevance threshold keeps the sampling budget from
being spent on failure modes that rarely govern.

\paragraph{2.4.3 Targeted space-filling
sampling}\label{targeted-space-filling-sampling}

A branch with too few training points does not provide a reliable fit,
so its \(U\) values cannot indicate where additional points are needed.
Among the candidate points at which the underpopulated branch governs,
the point farthest from the existing training points in scaled feature
space is therefore selected until the branch reaches the minimum
population. Afterwards, the \(U\)-criterion takes over. For the low load
ratios (\(\nu=0.1\), and \(\nu=0.2\) at \(\lambda=60\)), where this rule
repeatedly missed the small region of the minority mode in preliminary
studies, the candidate closest to an existing training point of the
branch is selected instead, and five space-filling points are added once
the minimum population is reached.

\paragraph{2.4.4 Isotropic kernel
fallback}\label{isotropic-kernel-fallback}

The Matérn kernel has three kinds of hyperparameters. The length scales
set how quickly the prediction may change along each input variable, the
kernel amplitude sets the magnitude of \(\hat\sigma\) far from the
training points, and the noise variance of the nugget term allows small
deviations from the training values. With one length scale per input
variable, the anisotropic kernel has \(d+2=7\) hyperparameters for
\(d=5\) inputs. With only 3--5 points, this fit is underdetermined. The
optimizer drives the noise variance to its lower bound and yields a
spuriously small amplitude and \(\hat\sigma\), to which the
\(U\)-criterion is directly sensitive. As an option, a branch below the
minimum population therefore uses an isotropic kernel with a single
length scale and three hyperparameters. This option was used only for
the low load ratios together with the variant of Section 2.4.3, not for
the examples of Section 4.

\paragraph{2.4.5 Classifier-assisted
win-share}\label{classifier-assisted-win-share}

A direct estimate of the win-share in Eq. (\ref{eq-winshare-def})
replaces the unknown governing mode \(c(\mathbf{x}_i)\) by the branch
that gives the smaller prediction in Eq. (\ref{eq-grenzfl}),

\begin{equation}\protect\phantomsection\label{eq-winshare}{\tilde s_b = \frac{1}{N_c}\sum_{i=1}^{N_c}\mathbb{1}\big[\hat g_b(\mathbf{x}_i) = \hat g(\mathbf{x}_i)\big],}\end{equation}

where \(\hat g_b\) is the prediction of branch \(b\). This estimate is
biased against data-scarce branches. Outside its bounding box, a branch
cannot govern, so a branch with few points and a small box is
underestimated even where its mode governs a larger part of the pool,
and it might never receive the points it needs. The win-share is
therefore estimated with a GP classifier that is trained only on the FE
labels and is not restricted to any box. It returns for every point
\(\mathbf{x}\) the probability \(p_b(\mathbf{x})\) that mode \(b\)
governs there, and

\begin{equation}\protect\phantomsection\label{eq-winshare-clf}{\hat s_b = \frac{1}{N_c}\sum_{i=1}^{N_c}\mathbb{1}\big[p_b(\mathbf{x}_i) \geq 0.5\big].}\end{equation}

Eqs. (\ref{eq-winshare-def}) to (\ref{eq-winshare-clf}) differ only in
how the governing mode at a pool point is decided, by the FE analysis
(known only at the training points), by the surrogate (biased by the
box) or by the classifier (used in this work). The classifier is
refitted at every iteration on all FE labels and evaluated on the
candidate pool of the \(U\)-criterion. It only decides whether a
data-scarce branch receives additional points and does not enter the
surrogate prediction.

\subsubsection{2.5 Model-uncertainty-aware active
learning}\label{model-uncertainty-aware-active-learning}

The surrogate is trained on the limit-state function of Eq.
(\ref{eq-limitstate}) with \(\Theta_R=\Theta_E=1\), because the FE model
provides only the resistance, and the model uncertainties have no
counterpart in it. They are nevertheless included in the \(U\)-criterion
during active learning. Since \(E=N_e\) is known at every candidate
point, the surrogate gives the resistance prediction
\(\hat R(\mathbf{x}) = \hat g(\mathbf{x}) + E\) with the same predictive
standard deviation \(\hat\sigma(\mathbf{x})\). With \(\Theta_R\) and
\(\Theta_E\) as factors,

\begin{equation}\protect\phantomsection\label{eq-theta-moments}{\hat g_\Theta(\mathbf{x}) = \Theta_R\,\hat R(\mathbf{x}) - \Theta_E\,E, \qquad \hat\sigma_\Theta(\mathbf{x}) = \Theta_R\,\hat\sigma(\mathbf{x}).}\end{equation}

The term \(\Theta_E E\) only shifts the prediction, so \(\Theta_E\)
enters the \(U\)-criterion through its numerator,

\begin{equation}\protect\phantomsection\label{eq-u-theta}{U = \frac{|\hat g_\Theta|}{\hat\sigma_\Theta} = \frac{|\Theta_R\,\hat R(\mathbf{x}) - \Theta_E\,E|}{\Theta_R\,\hat\sigma(\mathbf{x})}.}\end{equation}

\(\Theta_R\) and \(\Theta_E\) are drawn for every point of the candidate
pool, so active learning places FE evaluations near the failure boundary
including the model uncertainties. In preliminary campaigns without this
step, the FORM design point lay at concrete strengths up to about 8 MPa
below the range of the training points, where the surrogate only
extrapolated. A tighter convergence threshold did not help, because
over-smoothed length scales reported a falsely low \(\hat\sigma\) there.

\subsubsection{2.6 Post-convergence
refinement}\label{post-convergence-refinement}

In the two-branch formulation, convergence of the \(U\)-criterion is not
sufficient. The confidence interval on \(\beta\) can remain wide, and a
data-scarce branch can be miscalibrated without this being reflected in
\(U\). Two refinement steps therefore follow convergence. Unlike
existing accuracy measures of the failure-probability estimate {[}26{]},
they are triggered by an empirical calibration check and add points to
the branch that drives the uncertainty.

In the leave-one-out check, each training point \(\mathbf{x}_j\) of a
branch close to the limit state is removed in turn, and the branch is
refitted without it. The standardized residual is

\begin{equation}\protect\phantomsection\label{eq-loo}{z_j = \frac{y_j - \hat g_{-j}(\mathbf{x}_j)}{\hat\sigma_{-j}(\mathbf{x}_j)},}\end{equation}

where \(y_j = R_j - E_j\) is the limit-state value from the FE analysis
at \(\mathbf{x}_j\) with \(\Theta_R=\Theta_E=1\), and \(\hat g_{-j}\)
and \(\hat\sigma_{-j}\) are the prediction and the predictive standard
deviation of the branch refitted without \(\mathbf{x}_j\). For a
calibrated branch, \(z_j\) should lie roughly within \([-3, +3]\)
{[}27{]}. If \(\max|z_j| \geq 3\) and the branch governs at least 1\% of
the pool, the three candidate points closest to its worst point are
evaluated by FE, for at most three rounds.

If no branch is marked, the confidence-width follow-up evaluates
\(\Delta\beta\) (Section 3.1) on \(10^7\) pool points. If it exceeds a
target (0.1 in this study, a choice and not a code requirement), four
points per round are added with maximal distance to the existing
training points, in the region of the branch that contributes most of
the pool points failing only under the pessimistic shifted limit-state
function \(G^-\). The follow-up stops at the target, after 40 additional
points, or after three rounds without improvement. Where both modes are
nearly equally likely along the mode boundary, \(\Delta\beta\) decreases
slowly, so the target was reached only in part of the parameter study.
The remaining width is reported with \(\beta\). Because the added points
change the surrogate, \(\min U\) fell below \(u_{\text{threshold}}\)
again in 14 of 53 configurations. Active learning is then resumed with
the same stopping rule, the leave-one-out check is repeated, and no
further confidence-width follow-up is run.

\subsubsection{2.7 Overall procedure}\label{overall-procedure}

\begin{figure}[htbp]

\centering{

\includegraphics[width=0.6\linewidth,height=\textheight,keepaspectratio]{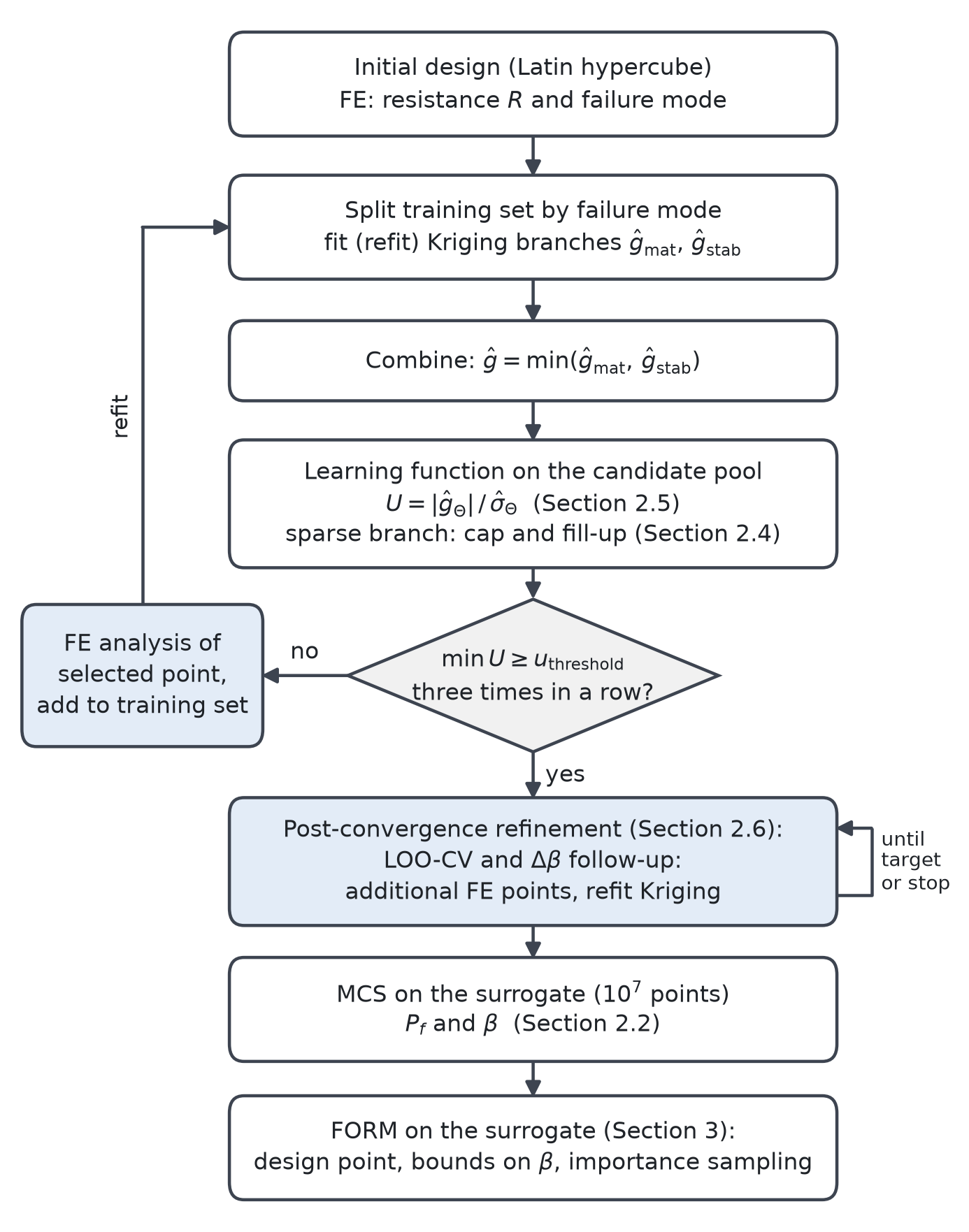}

}

\caption{\label{fig-ablauf}Overall procedure for piecewise (two-branch)
AK-MCS}

\end{figure}

Fig.~\ref{fig-ablauf} summarizes the procedure. The GP regressors are
implemented with scikit-learn {[}28{]}. With only a few training points,
the kernel fit of a branch is not unique. Refitting the same two
training points with different optimizer starting values gave up to an
eight-fold difference in \(\hat\sigma\), and the same training data gave
\(\min U = 0.014\) in one run and 13.6 in another. The loop therefore
stops only once \(\min U \geq u_{\text{threshold}}\) in three
consecutive iterations. The refinement of Section 2.6 follows, and
\(P_f\) and \(\beta\) are evaluated both by MCS and by the FORM search
on the same surrogate (Section 3).

\subsection{3 Confidence Bounds and FORM on the Two-Branch Kriging
Surface}\label{confidence-bounds-and-form-on-the-two-branch-kriging-surface}

\subsubsection{3.1 Confidence interval from the Kriging prediction
uncertainty}\label{confidence-interval-from-the-kriging-prediction-uncertainty}

The surrogate predicts \(\hat g(\mathbf{x})\) with
\(\Theta_R=\Theta_E=1\), which is equivalent to a resistance prediction
\(\hat R(\mathbf{x}) = \hat g(\mathbf{x}) + E\) with the same
\(\hat\sigma(\mathbf{x})\) (Section 2.5). At every pool point \(i\), the
resistance prediction is shifted by \(\pm\kappa\hat\sigma\) (two-sided
95\%, \(\kappa = 1.96\)) before the model uncertainties are applied as
in Eq. (\ref{eq-theta-moments}), giving the pessimistic and optimistic
shifted limit-state functions \(G_i^{\mp}\),

\begin{equation}\protect\phantomsection\label{eq-bounds}{G_i^{\mp} = \Theta_{R,i}\big(\hat R(\mathbf{x}_i) \mp \kappa\,\hat\sigma(\mathbf{x}_i)\big) - \Theta_{E,i}\,E_i = \hat g_\Theta(\mathbf{x}_i) \mp \kappa\,\hat\sigma_\Theta(\mathbf{x}_i)}\end{equation}

\(\Theta_R\) scales the whole shifted prediction and therefore also
appears in front of \(\hat\sigma\), while \(\Theta_E E\) only shifts it.
Counting failures over \(N\) pool points with the nominal prediction and
with each shifted limit-state function gives

\begin{equation}\protect\phantomsection\label{eq-pf-bounds}{P_f=\frac1N\sum_i\mathbb 1[\hat g_\Theta(\mathbf x_i)\le0],\quad
P_f^+=\frac1N\sum_i\mathbb 1[G_i^-\le0],\quad
P_f^-=\frac1N\sum_i\mathbb 1[G_i^+\le0],}\end{equation}

and \(\beta = -\Phi^{-1}(P_f)\) applied to each estimate yields
\(\beta_{\text{lower}} \leq \beta \leq \beta_{\text{upper}}\) and
\(\Delta\beta = \beta_{\text{upper}} - \beta_{\text{lower}}\).
\(\beta_{\text{MCS}}\) denotes the nominal value obtained by the Monte
Carlo simulation. Bounds of this kind, obtained from
\(\hat g \pm 2\hat\sigma\), also perform best as a stopping criterion in
active learning when accuracy is the priority {[}12{]}.

\(\Delta\beta\) differs from the sampling error of the Monte Carlo
estimate. The sampling error shrinks with more pool points, whereas
\(\Delta\beta\) reflects how much the surrogate's prediction could
plausibly be wrong and does not depend on the pool size. Both are
therefore reported separately. Since \(\Delta\beta\) describes the
uncertainty in both directions, each bound is a one-sided 97.5\% bound,
more conservative than the one-sided 95\% level of the draft DAfStb
guideline D375 {[}8{]}.

\subsubsection{3.2 Calibration check via leave-one-out
cross-validation}\label{calibration-check-via-leave-one-out-cross-validation}

The bounds of Section 3.1 are only meaningful if \(\hat\sigma\) is a
calibrated measure of the prediction error. The likelihood-based
hyperparameter fit does not guarantee this, in particular for branches
with few points (Section 2.4.4). Calibration is therefore checked by
leave-one-out cross-validation {[}27{]} with the residuals \(z_j\) of
Eq. (\ref{eq-loo}) at the (at most) 30 training points of each branch
closest to the limit state. The hyperparameters are re-optimized for
every left-out point. Jones et al. {[}27{]} keep them fixed for
efficiency, noting that this is justified unless there are very few
observations, which is the case for data-scarce branches. For a
calibrated branch, \(z_j\) follows a standard normal distribution, with
a mean absolute value of about 0.8 and about 5\% of the values exceeding
2 in magnitude. A branch is marked for refinement if
\(\max|z_j| \geq 3\) and it governs at least 1\% of the candidate pool.
For the two example configurations, \(\max|z_j|\) is 0.69/0.70 and
2.42/2.00 for the material/stability branches (Section 4.3). Section 4.4
shows a branch with \(\max|z_j| = 6.93\) that is not marked because it
governs only 0.28\% of the pool.

\subsubsection{3.3 Correlated versus independent surrogate error and the
use of the
bounds}\label{correlated-versus-independent-surrogate-error-and-the-use-of-the-bounds}

The bounds of Section 3.1 shift every pool point simultaneously by
\(\pm\kappa\hat\sigma\), which assumes a fully correlated surrogate
error, the most conservative assumption. The opposite extreme treats the
error as independent at each point,
\(R = \hat R(\mathbf{x}) + \varepsilon\) with
\(\varepsilon \sim N(0,\hat\sigma(\mathbf{x})^2)\). Since \(E\),
\(\Theta_R\) and \(\Theta_E\) are fixed at each pool point (Section
2.5), the expected failure probability then follows in closed form,
without additional samples,

\begin{equation}\protect\phantomsection\label{eq-pf-indep}{P_f = \frac{1}{N}\sum_i \Phi\left(-\frac{\hat g_\Theta(\mathbf{x}_i)}{\hat\sigma_\Theta(\mathbf{x}_i)}\right),}\end{equation}

with each term weighted by \(w(\mathbf{u}_i)\) if the pool is drawn by
importance sampling (Section 3.6).

Both treatments are evaluated on the same importance-sampling pool for
the two example configurations \texttt{l60n07} and \texttt{l60n09} of
Section 4.1, so the differences reflect only the error treatment. The
independent-error treatment shifts \(\beta\) by only
\(\delta\beta_{\text{indep}} = \beta-\beta_{\text{indep}} = 0.0044\) in
\texttt{l60n07} (\(\beta_{\text{indep}}=4.245\) against \(\beta=4.249\))
and 0.0041 in \texttt{l60n09} (\(\beta_{\text{indep}}=4.088\) against
\(\beta=4.092\)), about 1/35 of the confidence widths
\(\Delta\beta = 0.151\) and 0.153. Independent errors largely cancel
over the pool, because points near the limit state are pushed into
failure about as often as out of it. Under full correlation, all points
move together. The actual, partially correlated effect of the surrogate
error is therefore expected to lie between
\(\delta\beta_{\text{indep}}\) and \(\Delta\beta/2\).

The two quantities answer different questions.
\(\delta\beta_{\text{indep}}\) shows that the nominal \(\beta\) is not
biased by the surrogate error, so \(\beta\) serves as the best estimate
of the reliability index. A verification against a target reliability,
in contrast, requires a lower bound such as \(\beta_{\text{lower}}\),
which is conservative because it assumes fully correlated errors. Both
statements hold only if \(\hat\sigma\) is calibrated, as checked by the
leave-one-out residuals (Section 3.2) and the FE spot check (Section
4.3).

\subsubsection{3.4 FORM/MPP search on the two-branch Kriging limit-state
surface}\label{formmpp-search-on-the-two-branch-kriging-limit-state-surface}

In addition to the pool-based estimate, the design point is searched
directly on the surrogate
\(\hat g(\mathbf{u}) = \min(\hat g_{\text{mat}}(\mathbf{u}), \hat g_{\text{stab}}(\mathbf{u}))\)
in standard normal space, with the basic variables obtained from
\(\mathbf{u}\) by their marginal transformations. This point is also
called the most probable point (MPP), hence the subscript MPP. Since
\(\hat g\) is inexpensive, the constrained problem
\(\min\|\mathbf{u}\|\) subject to \(\hat g(\mathbf{u})=0\) is solved by
sequential least-squares quadratic programming (SLSQP) as implemented in
SciPy {[}29{]}, with finite-difference gradients, instead of the
Hasofer--Lind--Rackwitz--Fiessler iteration {[}30,31{]}. Because the
kink between the branches can trap gradient steps, the search starts
from \(2d+4\) points at radius 3, one in each positive and negative
coordinate direction and four in random directions. A candidate is
accepted if SLSQP converges and \(|\hat g(\mathbf{u}^*)| \le 0.01\)~kN.
The accepted candidate of minimum norm gives
\(\beta_{\text{MPP}} = \|\mathbf{u}^*\|\), taken as negative if the
origin lies in the failure domain. The same search applied to
\(\hat g_{\text{mat}}\) and \(\hat g_{\text{stab}}\) individually gives
each branch's own design point, used in Section 3.6.

Unlike the pool-based estimate, \(\beta_{\text{MPP}}\) does not depend
on failures in a finite pool and remains available where MCS gives
\(P_f = 0\). For \texttt{l60n07} and \texttt{l60n09},
\(\beta_{\text{MPP}}\) exceeds \(\beta_{\text{MCS}}\) by 0.05 and 0.04,
and an FE analysis at \(\mathbf{u}^*\) confirms the surrogate prediction
there (Section 4.3).

\subsubsection{3.5 FORM on the shifted limit-state
surfaces}\label{form-on-the-shifted-limit-state-surfaces}

Repeating the MPP search on the shifted limit-state surfaces \(G^-\) and
\(G^+\) of Eq. (\ref{eq-bounds}) gives \(\beta^-_{\text{MPP}}\) and
\(\beta^+_{\text{MPP}}\), and
\(\Delta\beta_{\text{FORM}} = \beta^+_{\text{MPP}} - \beta^-_{\text{MPP}}\).
Whereas \(\Delta\beta\) counts all pool points whose sign can change
anywhere along the limit state, \(\Delta\beta_{\text{FORM}}\) depends
only on \(\hat\sigma\) near the design point, where active learning
concentrates the training points. In the parameter study,
\(\Delta\beta_{\text{FORM}}\) was smaller than \(\Delta\beta\) in 69 of
73 configurations with \(P_f>0\) in the pool (median ratio 0.52). It
becomes large where the design point lies far from all training points,
as at high \(\beta\), where \(\hat\sigma\) at the design point can
exceed its value at the training points by more than two orders of
magnitude. \(\Delta\beta_{\text{FORM}}\) therefore does not replace
\(\Delta\beta\), but it provides an interval where the pool contains no
failure.

\subsubsection{3.6 Importance sampling at the FORM design
points}\label{importance-sampling-at-the-form-design-points}

The draft DAfStb guideline D375 accepts a simulation result if the
coefficient of variation of \(P_f\) is at most 2.5\% {[}8, Clause
5.3(2){]}. Direct MCS needs \(N \approx 1600/P_f\) samples for this,
more than \(10^{12}\) at \(\beta = 6\). Importance sampling draws from a
density \(q(\mathbf{u})\) centered near the limit-state surface and
weights each sample by
\(w(\mathbf{u}) = \varphi(\mathbf{u})/q(\mathbf{u})\), which keeps the
estimate unbiased. Here,
\(q(\mathbf{u}) = \frac{1}{K}\sum_{k=1}^{K}\varphi(\mathbf{u}-\mathbf{u}_k^*)\)
is an equally weighted mixture of standard normal densities centered at
up to four design points, the nominal and pessimistic
(\(\kappa = 1.645\)) design points of the combined surrogate and the
design point of each branch. Centers closer than 0.5 in standard normal
space are merged. Yang et al. {[}19{]} similarly center a mixture at
several MPPs of a system surrogate but find them by a multimodal search,
whereas here they follow directly from the branches. Sampling runs in
blocks of \(5\cdot10^4\) until the nominal and the pessimistic \(P_f\)
both reach a coefficient of variation of 2.5\%, estimated from the
spread of the block estimates (at least five blocks, at most
\(4\cdot10^6\) samples).

The mixture is fixed after the FORM search. AK-IS {[}32{]} likewise
fixes a single density at one design point and assumes that no secondary
failure region exists. A failure region far from all centers is then
under-sampled, and \(P_f\) is underestimated. The block-wise coefficient
of variation does not reveal this, because every block misses the region
in the same way, a limitation also found in the benchmark of Moustapha
et al. {[}12{]}. With two failure modes, this is the typical risk. In
both example configurations, the combined design point lies on the
stability side, yet the material branch accounts for a large share of
the pool failures (Section 4.3). The branch design points give every
mode its own samples, and all samples enter a single \(P_f\) estimate on
the combined surrogate, weighted by the full mixture density.

\subsection{4 Application to a Representative Column
Example}\label{application-to-a-representative-column-example}

\subsubsection{4.1 Column configuration}\label{column-configuration}

\begin{figure}[htbp]

\centering{

\includegraphics[width=0.5\linewidth,height=\textheight,keepaspectratio]{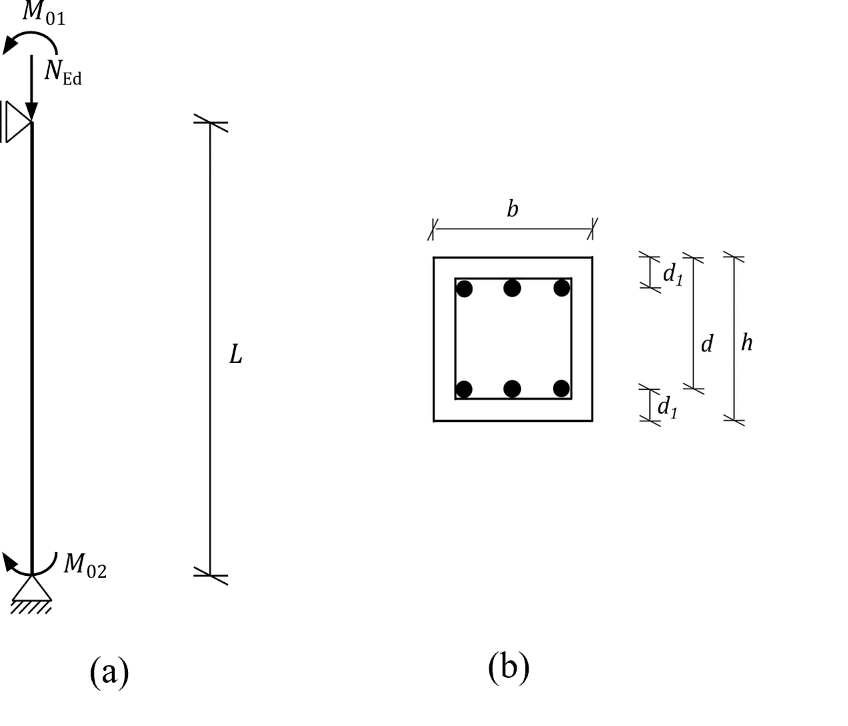}

}

\caption{\label{fig-column}(a) Simply supported column with equal
first-order end moments \(M_{01}=M_{02}\) and axial force \(N_{Ed}\);
(b) cross-section with six bars, three per layer (\(b\), \(h\), \(d\),
\(d_1\)).}

\end{figure}

The method is illustrated on the slender FRP-RC column example Omega-04
(Fig.~\ref{fig-column}). The column is pinned at the bottom and
vertically movable at the top, with a rectangular cross-section, six FRP
bars (three per layer) and concrete grade C50/60 according to EC2. It is
loaded by the design axial force \(N_{Ed}\) at the first-order
eccentricity \(e_0\) with the equal first-order end moments

\begin{equation}\protect\phantomsection\label{eq-endmoments}{M_{01}=M_{02}=N_{Ed}\,e_0 .}\end{equation}

Table~\ref{tbl-structural} lists the structural and material parameters.
The example is part of an ongoing parameter study over
\(\lambda \in \{40, 60, 80\}\), \(\nu = 0.1\) to 0.9 in steps of 0.1
(with an additional \(\nu = 0.85\) at \(\lambda = 80\)) and a stiffness
factor \(k_E \in \{0.9, 1.0, 1.1\}\) on the modulus of elasticity of the
concrete model of JCSS {[}1{]}, i.e.~84 configurations with
\(\text{CoV}(\Theta_R)=0.15\). Each configuration is labeled by its
slenderness ratio and axial load ratio, e.g.~\texttt{l60n07} for
\(\lambda = 60\) and \(\nu = 0.7\). Two configurations with mixed
failure are used here, \texttt{l60n07} and \texttt{l60n09}
(\(\lambda=60\), \(\nu=0.7\) and 0.9, \(k_E=1.0\)), with 68\% and 63\%
of their final training points failing in the material mode. A third
configuration, \texttt{l80n04} (\(\lambda=80\), \(\nu=0.4\)), is
dominated by stability and serves as a contrasting case. All slenderness
ratios considered exceed the limit \(\lambda_{\lim} \le 21.7\) for short
FRP-RC columns proposed in Annex O of prEN 1992-1-1:2023 {[}3{]} by a
factor of at least 1.8, so that stability failure may govern in any
configuration.

\begin{longtable}[]{@{}
  >{\raggedright\arraybackslash}p{(\linewidth - 4\tabcolsep) * \real{0.5278}}
  >{\raggedright\arraybackslash}p{(\linewidth - 4\tabcolsep) * \real{0.1389}}
  >{\raggedright\arraybackslash}p{(\linewidth - 4\tabcolsep) * \real{0.3333}}@{}}
\caption{Structural and material parameters of column example
Omega-04.}\label{tbl-structural}\tabularnewline
\toprule\noalign{}
\begin{minipage}[b]{\linewidth}\raggedright
Parameter
\end{minipage} & \begin{minipage}[b]{\linewidth}\raggedright
Symbol
\end{minipage} & \begin{minipage}[b]{\linewidth}\raggedright
Value
\end{minipage} \\
\midrule\noalign{}
\endfirsthead
\toprule\noalign{}
\begin{minipage}[b]{\linewidth}\raggedright
Parameter
\end{minipage} & \begin{minipage}[b]{\linewidth}\raggedright
Symbol
\end{minipage} & \begin{minipage}[b]{\linewidth}\raggedright
Value
\end{minipage} \\
\midrule\noalign{}
\endhead
\bottomrule\noalign{}
\endlastfoot
Cross-section (width \(\times\) depth) & \(b\times h\) & 300 \(\times\)
300 mm \\
Ratio of effective concrete cover to depth & \(d_1/h\) & 0.1 \\
Column length, \(L=\lambda\cdot i\), \(i=h/\sqrt{12}=86.6\) mm & \(L\) &
5196.6 mm (\(\lambda=60\)), 6928.8 mm (\(\lambda=80\)) \\
Slenderness ratio & \(\lambda\) & 60, 80 (examples) \\
Characteristic concrete compressive strength & \(f_{ck}\) & 50 MPa \\
Mean concrete compressive strength, \(f_{ck}+8\) & \(f_{cm}\) & 58
MPa \\
Coefficient for long-term effects on compressive strength &
\(\alpha_{cc}\) & 0.85 \\
Design concrete compressive strength, \(\alpha_{cc}f_{ck}/\gamma_c\) &
\(f_{cd}\) & 28.33 MPa \\
Modulus of elasticity of FRP bar & \(E_{\text{FRP}}\) & 40,000 MPa \\
Ultimate design strain of FRP bar & \(\varepsilon_{tud}\) & 10 mm/m \\
FRP reinforcement ratio & \(\rho_{\text{FRP}}\) & 0.0283 \\
\end{longtable}

\begin{figure}[htbp]

\centering{

\includegraphics[width=0.7\linewidth,height=\textheight,keepaspectratio]{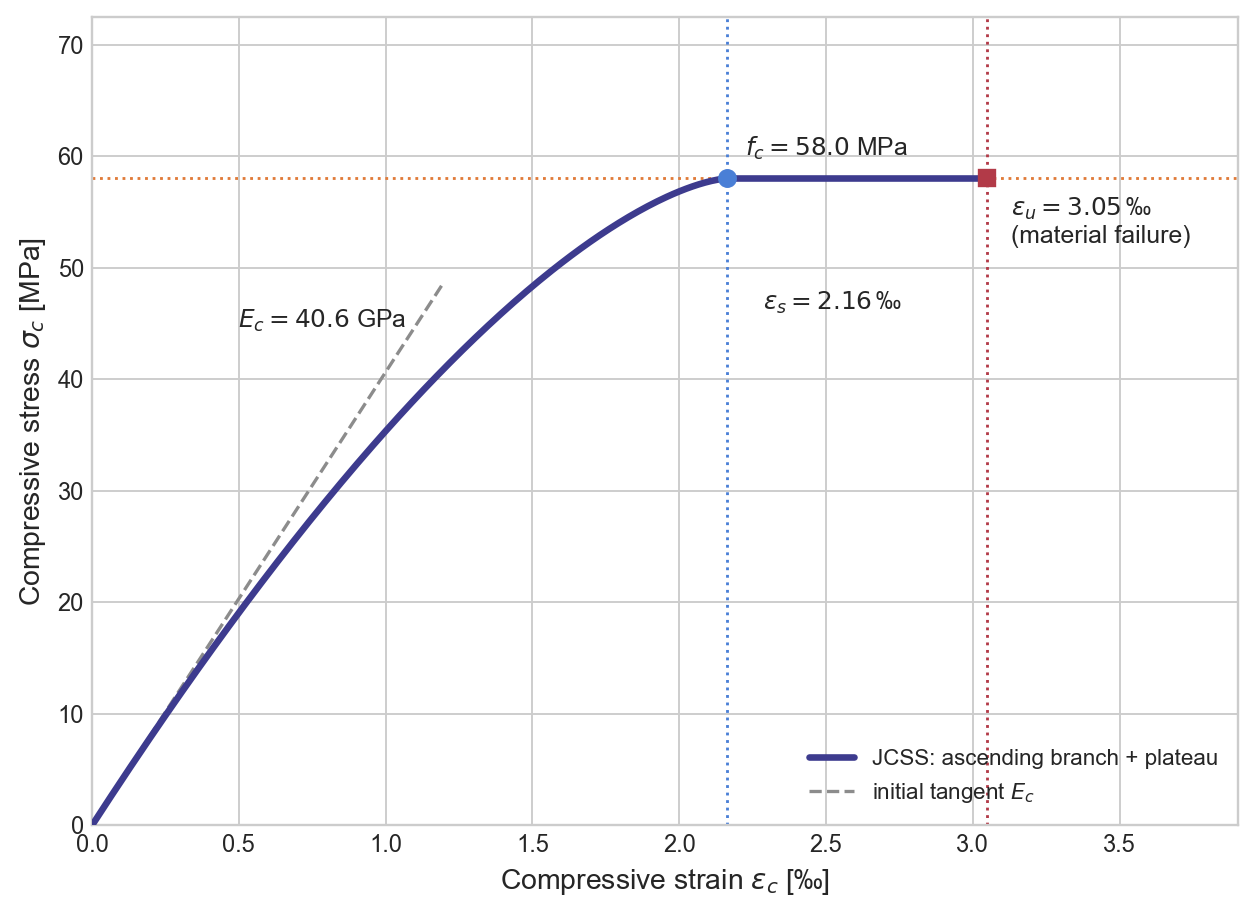}

}

\caption{\label{fig-concrete}JCSS concrete-in-compression law at mean
values (\(f_c=58\) MPa, \(Y_3=Y_4=1.0\), \(k_E=1.0\)).}

\end{figure}

The concrete stress--strain relation follows the JCSS Probabilistic
Model Code, Part 3.1 {[}1{]}, for short-term loading. The initial
tangent modulus is \(E_c = k_E\cdot10.5\,f_c^{1/3}Y_3\cdot1000\), where
\(k_E\) varies the stiffness independently of \(f_c\), as for different
aggregate types in EC2 (here \(k_E=1.0\)). With the reference strain
\(\varepsilon_s=0.0011\,f_c^{1/6}\), called \(\varepsilon_{c1}\) in EC2
{[}6{]}, and the shape factor \(k=E_c\varepsilon_s/f_c\), the ascending
branch is

\begin{equation}\protect\phantomsection\label{eq-jcssascending}{\sigma_c=f_c\left[1-(1-\eta)^{k}\right], \qquad \eta=\varepsilon/\varepsilon_s,}\end{equation}

followed by a plateau at \(f_c\) up to the ultimate strain
\(\varepsilon_u = 6\times10^{-3}f_c^{-1/6}Y_4\)
(Fig.~\ref{fig-concrete}). For any \(k>0\), the ascending branch reaches
\(f_c\) exactly at \(\varepsilon_s\) and never exceeds it. For \(k>1\)
it meets the plateau with zero slope, and for \(k=1\) with a kink that
is inherent in the JCSS formulation. For the rare samples with \(k<1\),
its curvature reverses. No floor on \(k\) is applied, consistent with
the literal JCSS formulation.

Table~\ref{tbl-randomvars} lists the basic random variables.
\(f_{cm,r}\) in Table~\ref{tbl-randomvars} is the cylinder strength,
termed \(f_{c0}\) in JCSS notation. JCSS Part 3.1 additionally reduces
it to the in-situ strength \(f_c=f_{c0}^{\lambda_{\text{JCSS}}}\,Y_1\)
with \(\lambda_{\text{JCSS}}=0.96\) and \(Y_1\) lognormal (mean 1.0, CoV
0.06). This reduction is not applied here. It would lower the mean
concrete strength by about 15\% and add further scatter.

\begin{longtable}[]{@{}
  >{\raggedright\arraybackslash}p{(\linewidth - 8\tabcolsep) * \real{0.4390}}
  >{\raggedright\arraybackslash}p{(\linewidth - 8\tabcolsep) * \real{0.1220}}
  >{\raggedleft\arraybackslash}p{(\linewidth - 8\tabcolsep) * \real{0.2073}}
  >{\raggedleft\arraybackslash}p{(\linewidth - 8\tabcolsep) * \real{0.0854}}
  >{\raggedright\arraybackslash}p{(\linewidth - 8\tabcolsep) * \real{0.1463}}@{}}
\caption{Basic random variables. }\label{tbl-randomvars}\tabularnewline
\toprule\noalign{}
\begin{minipage}[b]{\linewidth}\raggedright
Variable
\end{minipage} & \begin{minipage}[b]{\linewidth}\raggedright
Symbol
\end{minipage} & \begin{minipage}[b]{\linewidth}\raggedleft
Mean
\end{minipage} & \begin{minipage}[b]{\linewidth}\raggedleft
CoV
\end{minipage} & \begin{minipage}[b]{\linewidth}\raggedright
Distribution
\end{minipage} \\
\midrule\noalign{}
\endfirsthead
\toprule\noalign{}
\begin{minipage}[b]{\linewidth}\raggedright
Variable
\end{minipage} & \begin{minipage}[b]{\linewidth}\raggedright
Symbol
\end{minipage} & \begin{minipage}[b]{\linewidth}\raggedleft
Mean
\end{minipage} & \begin{minipage}[b]{\linewidth}\raggedleft
CoV
\end{minipage} & \begin{minipage}[b]{\linewidth}\raggedright
Distribution
\end{minipage} \\
\midrule\noalign{}
\endhead
\bottomrule\noalign{}
\endlastfoot
Concrete compressive strength & \(f_{cm,r}\) & 58 MPa & 0.15 &
Lognormal \\
Model uncertainty for \(E_c\) of concrete & \(Y_3\) & 1.0 & 0.15 &
Lognormal \\
Model uncertainty for ultimate strain \(\varepsilon_u\) & \(Y_4\) & 1.0
& 0.15 & Lognormal \\
Modulus of elasticity of FRP bar & \(E_{\text{FRP}}\) & 40,000 MPa &
0.05 & Lognormal \\
Dead load & \(G\) & \(G_m\) (Table~\ref{tbl-designload}) & 0.10 &
Normal \\
Live load & \(Q\) & \(Q_m\) (Table~\ref{tbl-designload}) & 0.26 &
Gumbel \\
Resistance model uncertainty & \(\Theta_R\) & 1.0 & 0.15 & Lognormal \\
Load model uncertainty & \(\Theta_E\) & 1.0 & 0.10 & Lognormal \\
\end{longtable}

The design axial force follows from the load ratio,

\begin{equation}\protect\phantomsection\label{eq-ned}{N_{Ed} = \nu\,A_c f_{cd},}\end{equation}

with the net concrete area \(A_c=A_g/(1+\rho_{\text{FRP}}) = 87{,}523\)
mm² and \(f_{cd}\) from Table~\ref{tbl-structural}. For the respective
\(\lambda\) and the mechanical reinforcement ratio
\(\omega' = E_{\text{FRP}}\rho_{\text{FRP}}/(100f_{cd}) = 0.4\), the
\(N\)--\(M\) interaction diagrams of Syed et al. {[}5{]} give the
dimensionless design moment \(\mu_{Ed}\), normalized consistently with
\(\nu\), and hence

\begin{equation}\protect\phantomsection\label{eq-e0h}{\frac{e_0}{h} = \frac{\mu_{Ed}}{\nu}.}\end{equation}

The characteristic loads are obtained by splitting \(N_{Ed}\) with the
partial factors \(\gamma_g=1.35\) and \(\gamma_q=1.5\) and a live-load
share \(\mu_q=0.2\) of the characteristic total load (\(\mu_g=0.8\)), as
in the reliability study of slender reinforced concrete columns by
Graubner and Six {[}9{]},

\begin{equation}\protect\phantomsection\label{eq-gkqk}{G_k = \frac{\mu_g\,N_{Ed}}{\gamma_g\mu_g+\gamma_q\mu_q},\qquad
Q_k = \frac{\mu_q\,N_{Ed}}{\gamma_g\mu_g+\gamma_q\mu_q}.}\end{equation}

This share lies at the lower end of the range 0.2--0.67 given for
typical building construction by Klein {[}33{]}. Following D375 {[}8{]},
the dead load is normally distributed with a coefficient of variation of
0.10 and \(G_m=G_k\), and the live load follows a Gumbel distribution
with a coefficient of variation of 0.26 and \(Q_m=Q_k/1.35\)
(Table~\ref{tbl-designload}).

\begingroup\footnotesize\setlength{\tabcolsep}{4pt}

\begin{longtable}[]{@{}
  >{\raggedright\arraybackslash}p{(\linewidth - 16\tabcolsep) * \real{0.1800}}
  >{\raggedleft\arraybackslash}p{(\linewidth - 16\tabcolsep) * \real{0.0600}}
  >{\raggedleft\arraybackslash}p{(\linewidth - 16\tabcolsep) * \real{0.0600}}
  >{\raggedleft\arraybackslash}p{(\linewidth - 16\tabcolsep) * \real{0.1100}}
  >{\raggedleft\arraybackslash}p{(\linewidth - 16\tabcolsep) * \real{0.0850}}
  >{\raggedleft\arraybackslash}p{(\linewidth - 16\tabcolsep) * \real{0.0850}}
  >{\raggedleft\arraybackslash}p{(\linewidth - 16\tabcolsep) * \real{0.1650}}
  >{\raggedleft\arraybackslash}p{(\linewidth - 16\tabcolsep) * \real{0.1100}}
  >{\raggedleft\arraybackslash}p{(\linewidth - 16\tabcolsep) * \real{0.1100}}@{}}
\caption{Design load, first-order end moments from Eq.
(\ref{eq-endmoments}) and back-calculated mean load values for the
example configurations. }\label{tbl-designload}\tabularnewline
\toprule\noalign{}
\begin{minipage}[b]{\linewidth}\raggedright
Configuration
\end{minipage} & \begin{minipage}[b]{\linewidth}\raggedleft
\(\lambda\)
\end{minipage} & \begin{minipage}[b]{\linewidth}\raggedleft
\(\nu\)
\end{minipage} & \begin{minipage}[b]{\linewidth}\raggedleft
\(N_{Ed}\)\\
{[}kN{]}\strut
\end{minipage} & \begin{minipage}[b]{\linewidth}\raggedleft
\(e_0/h\)
\end{minipage} & \begin{minipage}[b]{\linewidth}\raggedleft
\(e_0\)\\
{[}mm{]}\strut
\end{minipage} & \begin{minipage}[b]{\linewidth}\raggedleft
\(M_{01}=M_{02}\)\\
{[}kNm{]}\strut
\end{minipage} & \begin{minipage}[b]{\linewidth}\raggedleft
\(G_m\)\\
{[}kN{]}\strut
\end{minipage} & \begin{minipage}[b]{\linewidth}\raggedleft
\(Q_m\)\\
{[}kN{]}\strut
\end{minipage} \\
\midrule\noalign{}
\endfirsthead
\toprule\noalign{}
\begin{minipage}[b]{\linewidth}\raggedright
Configuration
\end{minipage} & \begin{minipage}[b]{\linewidth}\raggedleft
\(\lambda\)
\end{minipage} & \begin{minipage}[b]{\linewidth}\raggedleft
\(\nu\)
\end{minipage} & \begin{minipage}[b]{\linewidth}\raggedleft
\(N_{Ed}\)\\
{[}kN{]}\strut
\end{minipage} & \begin{minipage}[b]{\linewidth}\raggedleft
\(e_0/h\)
\end{minipage} & \begin{minipage}[b]{\linewidth}\raggedleft
\(e_0\)\\
{[}mm{]}\strut
\end{minipage} & \begin{minipage}[b]{\linewidth}\raggedleft
\(M_{01}=M_{02}\)\\
{[}kNm{]}\strut
\end{minipage} & \begin{minipage}[b]{\linewidth}\raggedleft
\(G_m\)\\
{[}kN{]}\strut
\end{minipage} & \begin{minipage}[b]{\linewidth}\raggedleft
\(Q_m\)\\
{[}kN{]}\strut
\end{minipage} \\
\midrule\noalign{}
\endhead
\bottomrule\noalign{}
\endlastfoot
\texttt{l60n07} & 60 & 0.7 & 1735.7 & 0.089 & 26.8 & 46.6 & 1006.2 &
186.3 \\
\texttt{l60n09} & 60 & 0.9 & 2231.6 & 0.028 & 8.5 & 19.0 & 1293.7 &
239.6 \\
\texttt{l80n04} & 80 & 0.4 & 991.8 & 0.072 & 21.5 & 21.4 & 575.0 &
106.5 \\
\end{longtable}

\endgroup

Each branch of the surrogate uses five input variables, \(f_c\),
\(E_c\), \(E_{\text{FRP}}\), \(N_e = G + Q\) and \(\varepsilon_u\),
which are the quantities the FE model actually receives. All points are
drawn in terms of the six basic variables \(f_c\), \(Y_3\), \(Y_4\),
\(E_{\text{FRP}}\), \(G\) and \(Q\) and mapped to these inputs.
\(\Theta_R\) and \(\Theta_E\) are not surrogate inputs, since they are
applied analytically (Section 2.5).

\subsubsection{4.2 Finite-element model and failure-mode
classification}\label{finite-element-model-and-failure-mode-classification}

Training data are generated with OpenSeesPy {[}34,35{]}, version
3.8.0.0. Each column is idealized as a two-dimensional fiber beam-column
with two force-based elements (\texttt{forceBeamColumn}), which satisfy
equilibrium exactly along each element. In preliminary studies with a
concrete law with a descending branch, this element became
ill-conditioned for low slenderness ratios and large eccentricities.
With the concrete model of JCSS used here, which has no descending
branch, it was sufficient for all configurations. Each section is
discretized into a \(30\times30\) grid of concrete fibers plus six FRP
fibers. The equilibrium path is traced by arc-length (Crisfield-type)
continuation, so that material softening and stability failure are
captured without a separate buckling analysis.

At every load step, the strains at all integration points are checked
against the strain limits of EN 1992-1-1 {[}6{]}, with the material
values of Section 4.1 in place of the design values. The strain at the
most compressed fiber is limited to \(\varepsilon_u\) (instead of
\(\varepsilon_{cu2}\)). In a fully compressed section, the strain at the
depth \(z_c = (1-\varepsilon_s/\varepsilon_u)\,h\) from the most
compressed fiber is limited to \(\varepsilon_s\) (instead of
\(\varepsilon_{c2}\) at \(3h/7\)). The tensile strain of the FRP bars is
limited to the rupture strain \(\varepsilon_{tud}\). If one of these
limits is reached (within 0.05‰) before the peak of the equilibrium
path, the failure is classified as \textbf{material failure}, and
\(N_R\) is the load at which the limit is first reached, interpolated
between load steps. Otherwise, \(N_R\) is the peak load, and the failure
is classified as \textbf{stability failure}. This label routes each FE
evaluation, with its single resistance value, to one branch (Section
2.3).

The material branch combines concrete crushing and FRP rupture. Treating
both as one smooth branch requires that at most one of them governs
across the relevant part of the pool. This holds for \texttt{l60n07} and
\texttt{l60n09}, where every material-mode training point reaches the
concrete limit and the FRP strain stays well below
\(\varepsilon_{tud}=10\) mm/m. In the parameter study, the condition
fails at \(\nu=0.1\), where 30--45\% of the material-mode training
points (\(k_E=1.0\)) reach the FRP limit first. These three
configurations are therefore excluded from the examples. They may
require a third branch, which is investigated in an ongoing study.

\subsubsection{4.3 Results}\label{results}

The data-scarcity safeguards of Section 2.4 were active in the two
mixed-failure configurations \texttt{l60n07} and \texttt{l60n09}. Five
(\texttt{l60n07}) and seven (\texttt{l60n09}) of the 15 initial points
fell into the stability branch, so the minimum-population rule chose the
next training point in the first 3 of 8 and the first 8 of 14
active-learning iterations. The win-shares of both branches remained
between 20\% and 80\%, so the classifier did not alter any decision. Of
the final 71 (\texttt{l60n07}) and 56 (\texttt{l60n09}) training points,
15 come from the initial design, 13 and 13 from active learning, 3 and 0
from the leave-one-out follow-up and 40 and 28 from the confidence-width
follow-up, which used its full budget in \texttt{l60n07}. Convergence
held after the follow-up (\(\min U = 5.69\) and 6.92). The final
\(\Delta\beta\) is 0.128 and 0.147, above the target of 0.1, and
\(\max|z_j|\) stays below 3 in both branches (Table~\ref{tbl-formmcs}).

Table~\ref{tbl-formmcs} presents the results of the FORM/MCS consistency
and leave-one-out check for the three illustrative configurations in the
final state with all training points. For \texttt{l60n07} and
\texttt{l60n09}, \(\beta_{\text{MPP}}\) is 0.047 and 0.036 (1.1\% and
0.9\%) larger than \(\beta_{\text{MCS}}\). FORM linearizes the limit
state at a single design point, which the surrogate places on the
stability side for both configurations (\(z_{\text{mode}}=0.41\) and
0.98, defined below). For \texttt{l60n07}, an FE analysis at this point
(Table~\ref{tbl-fespotcheck}) classifies it as material-governed
instead, a disagreement consistent with the surrogate's uncertainty this
close to the boundary. The other mode still contributes 48 of the 118
and 23 of the 228 pool failures, which motivates the branch centers of
Section 3.6. All leave-one-out residuals remain below 3.

\begingroup\scriptsize\setlength{\tabcolsep}{3pt}

\begin{longtable}[]{@{}
  >{\raggedright\arraybackslash}p{(\linewidth - 16\tabcolsep) * \real{0.1558}}
  >{\raggedleft\arraybackslash}p{(\linewidth - 16\tabcolsep) * \real{0.0390}}
  >{\raggedleft\arraybackslash}p{(\linewidth - 16\tabcolsep) * \real{0.0390}}
  >{\raggedright\arraybackslash}p{(\linewidth - 16\tabcolsep) * \real{0.1396}}
  >{\raggedright\arraybackslash}p{(\linewidth - 16\tabcolsep) * \real{0.1753}}
  >{\raggedleft\arraybackslash}p{(\linewidth - 16\tabcolsep) * \real{0.0747}}
  >{\raggedleft\arraybackslash}p{(\linewidth - 16\tabcolsep) * \real{0.0747}}
  >{\raggedleft\arraybackslash}p{(\linewidth - 16\tabcolsep) * \real{0.1461}}
  >{\raggedleft\arraybackslash}p{(\linewidth - 16\tabcolsep) * \real{0.1331}}@{}}
\caption{Results of the FORM/MCS consistency and leave-one-out check for
three illustrative configurations in the final state. n/a means no
failure in the pool. }\label{tbl-formmcs}\tabularnewline
\toprule\noalign{}
\begin{minipage}[b]{\linewidth}\raggedright
Configuration
\end{minipage} & \begin{minipage}[b]{\linewidth}\raggedleft
\(\lambda\)
\end{minipage} & \begin{minipage}[b]{\linewidth}\raggedleft
\(\nu\)
\end{minipage} & \begin{minipage}[b]{\linewidth}\raggedright
FE analyses\\
(mat / stab)\strut
\end{minipage} & \begin{minipage}[b]{\linewidth}\raggedright
Failures of \(10^7\)\\
(mat / stab)\strut
\end{minipage} & \begin{minipage}[b]{\linewidth}\raggedleft
\(\beta_{\text{MCS}}\)
\end{minipage} & \begin{minipage}[b]{\linewidth}\raggedleft
\(\beta_{\text{MPP}}\)
\end{minipage} & \begin{minipage}[b]{\linewidth}\raggedleft
\(\beta_{\text{IS}}\)\\
(CoV)\strut
\end{minipage} & \begin{minipage}[b]{\linewidth}\raggedleft
\(\max\lvert z_j\rvert\)\\
(mat / stab)\strut
\end{minipage} \\
\midrule\noalign{}
\endfirsthead
\toprule\noalign{}
\begin{minipage}[b]{\linewidth}\raggedright
Configuration
\end{minipage} & \begin{minipage}[b]{\linewidth}\raggedleft
\(\lambda\)
\end{minipage} & \begin{minipage}[b]{\linewidth}\raggedleft
\(\nu\)
\end{minipage} & \begin{minipage}[b]{\linewidth}\raggedright
FE analyses\\
(mat / stab)\strut
\end{minipage} & \begin{minipage}[b]{\linewidth}\raggedright
Failures of \(10^7\)\\
(mat / stab)\strut
\end{minipage} & \begin{minipage}[b]{\linewidth}\raggedleft
\(\beta_{\text{MCS}}\)
\end{minipage} & \begin{minipage}[b]{\linewidth}\raggedleft
\(\beta_{\text{MPP}}\)
\end{minipage} & \begin{minipage}[b]{\linewidth}\raggedleft
\(\beta_{\text{IS}}\)\\
(CoV)\strut
\end{minipage} & \begin{minipage}[b]{\linewidth}\raggedleft
\(\max\lvert z_j\rvert\)\\
(mat / stab)\strut
\end{minipage} \\
\midrule\noalign{}
\endhead
\bottomrule\noalign{}
\endlastfoot
\texttt{l60n07} & 60 & 0.7 & 71 (48 / 23) & 118 (48 / 70) & 4.228 &
4.275 & 4.249 (0.5\%) & 0.69 / 0.70 \\
\texttt{l60n09} & 60 & 0.9 & 56 (35 / 21) & 228 (23 / 205) & 4.077 &
4.113 & 4.092 (0.4\%) & 2.42 / 2.00 \\
\texttt{l80n04} & 80 & 0.4 & 47 (12 / 35) & 0 & n/a & 6.194 & 6.187
(0.8\%) & 2.16 / 0.97 \\
\end{longtable}

\endgroup

With 118 and 228 failures, the coefficient of variation of \(P_f\) is
9\% for \texttt{l60n07} and 7\% for \texttt{l60n09}, above the 2.5\%
required by D375. Importance sampling on the surrogate with the mixture
density of Section 3.6 meets this criterion with \(2.5\cdot10^5\)
samples. It gives \(\beta_{\text{IS}} = 4.249\) and 4.092, within about
one standard deviation of the direct estimates, with a coefficient of
variation of \(P_f\) of at most 0.5\% (in parentheses in
Table~\ref{tbl-formmcs}). For \texttt{l80n04}, where
\(\beta_{\text{MPP}}\approx6.2\) lies beyond the resolution of \(10^7\)
direct samples, it gives \(\beta_{\text{IS}} = 6.187\) (0.8\%). Because
the direct pool contains no failure, \(\Delta\beta\) is not defined
there. On the importance-sampling pool, \(\beta_{\text{lower}} = 6.08\)
and \(\beta_{\text{upper}} = 6.28\), so \(\Delta\beta = 0.21\), wider
than for \texttt{l60n07} and \texttt{l60n09} because the design point
lies farther from the training points. Although stability dominates,
material failure still governs 8\% of the pool, and the surrogate is
well calibrated, so the missing pool failures reflect the rarity of the
event, not a deficiency of the surrogate. A single FE analysis at the
combined design point checks the surrogate where it matters most for
\(\beta\). The FE resistance and the Kriging prediction differ by at
most 0.58\% (Table~\ref{tbl-fespotcheck}). The accuracy criteria of D375
are not applied systematically here, which is the subject of an ongoing
study.

\begin{longtable}[]{@{}
  >{\raggedright\arraybackslash}p{(\linewidth - 10\tabcolsep) * \real{0.2319}}
  >{\raggedleft\arraybackslash}p{(\linewidth - 10\tabcolsep) * \real{0.0870}}
  >{\raggedleft\arraybackslash}p{(\linewidth - 10\tabcolsep) * \real{0.0870}}
  >{\raggedleft\arraybackslash}p{(\linewidth - 10\tabcolsep) * \real{0.1884}}
  >{\raggedleft\arraybackslash}p{(\linewidth - 10\tabcolsep) * \real{0.1884}}
  >{\raggedleft\arraybackslash}p{(\linewidth - 10\tabcolsep) * \real{0.1884}}@{}}
\caption{FE spot check at the FORM design
point.}\label{tbl-fespotcheck}\tabularnewline
\toprule\noalign{}
\begin{minipage}[b]{\linewidth}\raggedright
Configuration
\end{minipage} & \begin{minipage}[b]{\linewidth}\raggedleft
\(\lambda\)
\end{minipage} & \begin{minipage}[b]{\linewidth}\raggedleft
\(\nu\)
\end{minipage} & \begin{minipage}[b]{\linewidth}\raggedleft
\(R_{\text{Kriging}}\)\\
{[}kN{]}\strut
\end{minipage} & \begin{minipage}[b]{\linewidth}\raggedleft
\(R_{\text{FE}}\)\\
{[}kN{]}\strut
\end{minipage} & \begin{minipage}[b]{\linewidth}\raggedleft
Relative\\
deviation\strut
\end{minipage} \\
\midrule\noalign{}
\endfirsthead
\toprule\noalign{}
\begin{minipage}[b]{\linewidth}\raggedright
Configuration
\end{minipage} & \begin{minipage}[b]{\linewidth}\raggedleft
\(\lambda\)
\end{minipage} & \begin{minipage}[b]{\linewidth}\raggedleft
\(\nu\)
\end{minipage} & \begin{minipage}[b]{\linewidth}\raggedleft
\(R_{\text{Kriging}}\)\\
{[}kN{]}\strut
\end{minipage} & \begin{minipage}[b]{\linewidth}\raggedleft
\(R_{\text{FE}}\)\\
{[}kN{]}\strut
\end{minipage} & \begin{minipage}[b]{\linewidth}\raggedleft
Relative\\
deviation\strut
\end{minipage} \\
\midrule\noalign{}
\endhead
\bottomrule\noalign{}
\endlastfoot
\texttt{l60n07} & 60 & 0.7 & 2455.3 & 2458.3 & 0.12\% \\
\texttt{l60n09} & 60 & 0.9 & 3028.2 & 3029.2 & 0.04\% \\
\texttt{l80n04} & 80 & 0.4 & 2010.5 & 1998.8 & \(-0.58\)\% \\
\end{longtable}

\begin{figure}[htbp]

\centering{

\pandocbounded{\includegraphics[keepaspectratio]{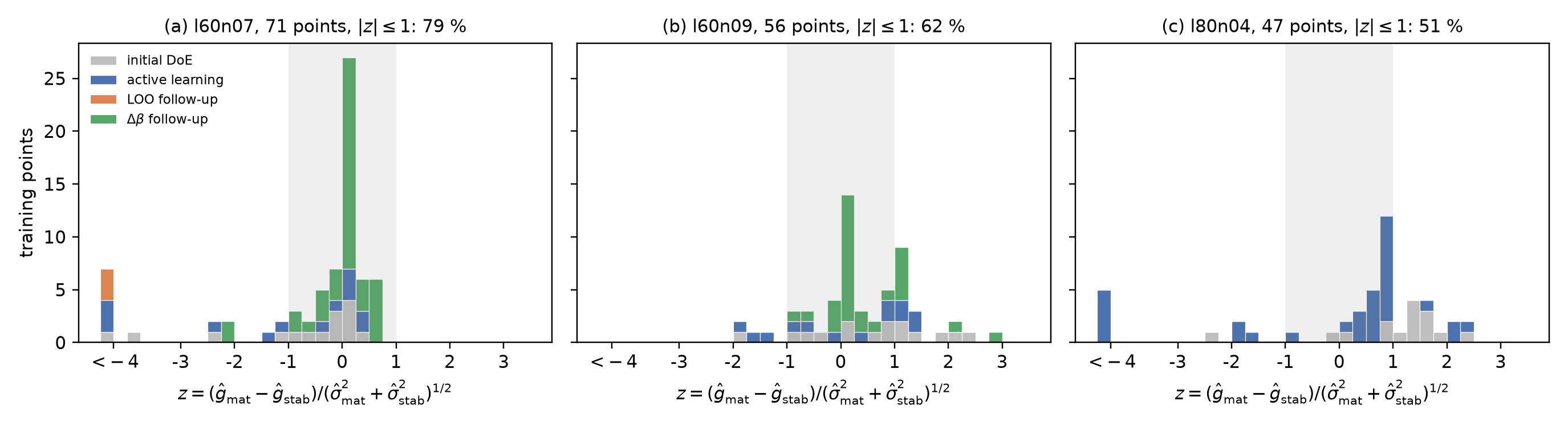}}

}

\caption{\label{fig-zdist}Normalized distance \(z_{\text{mode}}\) of the
training points to the mode boundary, by origin, for (a) \texttt{l60n07}
(71 points), (b) \texttt{l60n09} (56 points) and (c) \texttt{l80n04} (47
points) in the final state. The shaded band marks
\(|z_{\text{mode}}|\le1\).}

\end{figure}

Fig.~\ref{fig-zdist} shows where the training points lie relative to the
mode boundary. At each training point, the difference
\(\hat g_{\text{mat}}-\hat g_{\text{stab}}\) indicates the governing
mode (negative for material, positive for stability failure). Dividing
it by the combined predictive standard deviation gives the normalized
distance
\(z_{\text{mode}}=(\hat g_{\text{mat}}-\hat g_{\text{stab}})/(\hat\sigma_{\text{mat}}^2+\hat\sigma_{\text{stab}}^2)^{1/2}\).
For \(|z_{\text{mode}}|\le1\), the surrogate cannot decide which mode
governs. In \texttt{l60n07} and \texttt{l60n09}, 79\% and 62\% of the
training points lie in this range, and over all 60 configurations of the
parameter study in which both modes occur, the median share is 65\% and
the maximum 93\%. Most FE analyses are thus spent where both modes are
almost equally strong, exactly where a single Kriging model would have
to reproduce the kink of the minimum. Even in \texttt{l80n04}, where
stability governs 92\% of the pool, 51\% of the points lie near the
boundary and only 30\% clearly on the stability side, with the design
point just beyond this range (\(z_{\text{mode}}=1.75\)).

The split of the FE analyses between the failure modes therefore differs
from the split of the pool failures (Table~\ref{tbl-formmcs}). In
\texttt{l60n07}, 48 of the 71 analyses end in material failure, but the
material branch governs only 48 of the 118 failures. In \texttt{l60n09},
the numbers are 35 of 56 analyses and 23 of 228 failures. The share of
each branch in \(P_f\) is thus a result of the analysis, not an
assumption.

\subsubsection{4.4 Single-branch fallback and an irrelevant,
miscalibrated
branch}\label{single-branch-fallback-and-an-irrelevant-miscalibrated-branch}

Two mechanisms of the method take effect only when one mode is nearly
absent, which is common in the parameter study. They are illustrated by
\texttt{l40n04} (\(\lambda=40\), \(\nu=0.4\)), in which material failure
governs throughout, and \texttt{l40n07} (\(\lambda=40\), \(\nu=0.7\)),
the clearest of six configurations with an irrelevant, miscalibrated
branch (Table~\ref{tbl-singlemode}).

\begingroup\scriptsize\setlength{\tabcolsep}{2.5pt}

\begin{longtable}[]{@{}lrrrrrrrr@{}}
\caption{Two configurations dominated by one failure mode in the final
state, with training points, win-share and \(\max|z_j|\) per branch and
the Monte Carlo result (n/a means that the branch is not
fitted).}\label{tbl-singlemode}\tabularnewline
\toprule\noalign{}
Configuration & \(\lambda\) & \(\nu\) & \(k_E\) & Points (mat / stab) &
Win-share {[}\%{]} (mat / stab) & \(\max\lvert z_j\rvert\) (mat / stab)
& Failures of \(10^7\) & \(\beta_{\text{MCS}}\) \\
\midrule\noalign{}
\endfirsthead
\toprule\noalign{}
Configuration & \(\lambda\) & \(\nu\) & \(k_E\) & Points (mat / stab) &
Win-share {[}\%{]} (mat / stab) & \(\max\lvert z_j\rvert\) (mat / stab)
& Failures of \(10^7\) & \(\beta_{\text{MCS}}\) \\
\midrule\noalign{}
\endhead
\bottomrule\noalign{}
\endlastfoot
\texttt{l40n04} & 40 & 0.4 & 1.0 & 38 / 0 & 100 / n/a & 2.68 / n/a & 264
& 4.043 \\
\texttt{l40n07} & 40 & 0.7 & 1.0 & 43 / 4 & 99.72 / 0.28 & 1.67 / 6.93 &
158 & 4.162 \\
\end{longtable}

\endgroup

In \texttt{l40n04}, all 38 FE analyses end in material failure. The
stability branch is not fitted, and the prediction reduces to the
material branch without further intervention. Of the 38 training points,
15 come from the initial design, 7 from active learning and 16 from the
confidence-width follow-up. With 264 pool failures
(\(\text{CoV}(P_f)=6.2\)\%), a formal verification would still require
importance sampling. In \texttt{l40n07}, the stability branch has four
training points and \(\max|z_j|=6.93\), far above 3, but it governs only
0.28\% of the pool and is therefore not marked. None of the 158 pool
failures is governed by it, and the material branch alone gives the same
158 failures and \(\beta=4.162\). Refining the stability branch would
cost FE analyses without changing \(\beta\).

\subsection{5 Discussion}\label{discussion}

The two-branch formulation is limited to two failure modes. At
\(\nu=0.1\), concrete crushing and FRP rupture compete throughout the
pool (Section 4.2), and a third branch should be tested there. The
minimum combination and the \(U\)-criterion extend directly to more
modes, but each branch then receives fewer points, so data starvation
becomes more severe. The axis-aligned bounding box of Section 2.4.1 does
not prevent extrapolation in its empty corners.

Whether a single Kriging surrogate without branch-splitting would give
the same reliability index and the same local calibration is
investigated in a companion paper in preparation.

Over the 84 configurations, the final state required a median of 63 FE
analyses (range 17 to 172), 35\% of them from the post-convergence
refinement. With about one second per analysis, this is uncritical here,
but the method is intended for FE models with far longer runs.

Controlling the final accuracy by bounds on \(\beta\) rather than by the
learning function agrees with the benchmark of Moustapha et al.
{[}12{]}, in which such criteria gave the most accurate estimates. The
same benchmark found PC-Kriging {[}36{]} superior to ordinary Kriging,
mainly for problems with many input variables. With five inputs, only a
small benefit is expected here, but PC-Kriging could replace ordinary
Kriging within each branch without changing the framework. The fixed
importance-sampling density of Section 3.6 cannot detect a failure
region far from all centers. Adaptive densities built from the surrogate
itself {[}37{]} would distribute the samples over all failure regions
that the surrogate represents.

\subsection{6 Conclusions}\label{conclusions}

A piecewise, two-branch AK-MCS method was developed for slender FRP-RC
columns at medium to high axial load levels, for which a single FE
analysis returns only the capacity of the governing failure mode and the
transition between material and stability failure is broad. The main
findings for the column example are as follows.

\begin{enumerate}
\def\labelenumi{\arabic{enumi}.}
\tightlist
\item
  In configurations with mixed failure, 79\% and 62\% of the training
  points lie within one predictive standard deviation of the mode
  boundary. Separate branches combined by the minimum avoid fitting the
  kink in a region that contains most of the training points.
\item
  The safeguards against data starvation keep the method operational
  when one mode is rare or absent. A miscalibrated branch that governs
  less than 1\% of the candidate pool is not refined, because it does
  not affect \(\beta\).
\item
  Convergence of the \(U\)-criterion alone is not sufficient. The
  accuracy of \(\beta\) and of its lower bound \(\beta_{\text{lower}}\)
  is controlled by the post-convergence refinement, based on the
  leave-one-out check and on the width \(\Delta\beta\) of the confidence
  interval.
\item
  FORM on the surrogate agrees with the Monte Carlo estimate to within
  0.05 and provides \(\beta\) where the pool contains no failure. Where
  both modes contribute to the failures, it overestimates \(\beta\)
  slightly, because it represents only one failure region. An FE
  analysis at the design point confirmed the surrogate to within 0.6\%.
\item
  FORM on the shifted limit-state surfaces gives a confidence interval
  local to the design point, narrower than \(\Delta\beta\) in 69 of 73
  configurations. Importance sampling at the design points gave
  \(\beta = 6.19\) with a coefficient of variation of \(P_f\) of 0.8\%
  from \(2.5\cdot10^5\) samples where direct MCS with \(10^7\) samples
  found no failure.
\end{enumerate}

Whether these results generalize across the full design space, and
whether the method can replace an FE-based Level-III analysis in
accordance with D375 and EN 1990, remains the subject of an ongoing
study.

\subsection*{CRediT authorship contribution
statement}\label{credit-authorship-contribution-statement}
\addcontentsline{toc}{subsection}{CRediT authorship contribution
statement}

\textbf{Arndt Goldack:} Conceptualization, Methodology, Software,
Validation, Formal analysis, Investigation, Data curation,
Visualization, Writing -- original draft, Writing -- review and editing,
Supervision, Project administration. \textbf{Richard Koehler:}
Resources, Validation, Writing -- review and editing. \textbf{Ahmed
Usama Syed:} Resources, Validation, Writing -- review and editing.

\subsection*{Funding}\label{funding}
\addcontentsline{toc}{subsection}{Funding}

This research did not receive any specific grant from funding agencies
in the public, commercial, or not-for-profit sectors.

\subsection*{Declaration of competing
interest}\label{declaration-of-competing-interest}
\addcontentsline{toc}{subsection}{Declaration of competing interest}

The authors declare that they have no known competing financial
interests or personal relationships that could have appeared to
influence the work reported in this paper.

\subsection*{Declaration of generative AI and AI-assisted technologies
in the manuscript preparation
process}\label{declaration-of-generative-ai-and-ai-assisted-technologies-in-the-manuscript-preparation-process}
\addcontentsline{toc}{subsection}{Declaration of generative AI and
AI-assisted technologies in the manuscript preparation process}

During the preparation of this work the authors used Claude (Anthropic)
to assist with code development, data evaluation and language editing.
After using this tool, the authors reviewed and edited the content as
needed and take full responsibility for the content of the publication.

\subsection*{Data availability}\label{data-availability}
\addcontentsline{toc}{subsection}{Data availability}

Data will be made available on request.

\subsection*{References}\label{references}
\addcontentsline{toc}{subsection}{References}

\protect\phantomsection\label{refs}

\end{document}